\documentclass[12pt,preprint,flushrt]{aastex}
\usepackage{epsfig}
\usepackage{colordvi}
\usepackage{html}

\newcommand{\HI}{\mbox{{\rm H \footnotesize{I} }}}

\newcommand{\rtp}[1]{\ensuremath{^{#1}}}
\newcommand{\abs}[1]{\ensuremath{\mid #1 \mid}}
\newcommand{\apx}{\ensuremath{\sim}}
\newcommand{\pmt}{\ensuremath{\pm}}

\newcommand{\Vsysbar}{\ensuremath{\overline{V}_{sys}}}
\newcommand{\Vr}{\ensuremath{V_{r}}}
\newcommand{\Vyp}{\ensuremath{V_{y'}}}
\newcommand{\Vx}{\ensuremath{V_{x}}}
\newcommand{\Myp}{\ensuremath{\mu_{y'}}}
\newcommand{\Mx}{\ensuremath{\mu_{x}}}
\newcommand{\MypSQ}{\ensuremath{\mu^2_{y'}}}
\newcommand{\MxSQ}{\ensuremath{\mu^2_{x}}}
\newcommand{\Veib}{\ensuremath{\overline{V}_{ei}}}
\newcommand{\Vevb}{\ensuremath{\overline{V}_{ev}}}

\newcommand{\DPP }{\ensuremath{D_{PP}}}
\newcommand{\mM  }{\ensuremath{mM}}
\newcommand{\PP  }{\ensuremath{PP}}
\newcommand{\ppm }{\ensuremath{\mu}}
\newcommand{\Axr }{\ensuremath{\alpha_{x,r}}}
\newcommand{\Axyp}{\ensuremath{\alpha_{x,yp}}}
\newcommand{\Ayp }{\ensuremath{\alpha_{y'}}}

\newcommand{\muas}{\ensuremath{\mu\mbox{as}}}

\newcommand{\muasyr}{\ensuremath{\mu\mbox{as\,yr}^{-1}}}
\newcommand{\masyr}{\ensuremath{\mbox{mas} \; \mbox{yr}^{-1}}}
\newcommand{\kms}{\ensuremath{\mbox{km\,s}^{-1}}}

\newcommand{\etal}{{\em et al.}}
\begin{document}


\title{Galaxy Distances via Rotational Parallaxes}

\author{Rob P. Olling}
\affil{Rutgers University, NJ}
   \and
\author{D. M. Peterson}
\affil{SUNY at Stony Brook, NY}

\newpage

\begin{abstract}
   Astrometry is one of the foundations of astrophysics.  Accurate
distances to stars and galaxies allow for significant tests of stellar
evolution, galaxy formation and evolution and cosmology.  NASA's Space
Interferometry Mission (SIM) will obtain extraordinary precision
astrometry [1-10 \muas\ positional accuracy] for ten to twenty thousand
objects brighter than V=20 mag.
   In this paper we discuss a method to determine the distance of
nearby spiral galaxies using the technique of rotational
parallaxes. We show that it is possible to achieve distance errors of
only a few percent using this method. With such distances at hand, it
becomes possible to determine an accurate zero-point for the
Tully-Fisher relation, one of the tools to measure distances
throughout the universe. Precision cosmology will become possible once
the Hubble constant has been accurately determined from a SIM-based
calibration of the Tully-Fisher relation.
   The rotational parallax method employs the common motions of a
number of stars to determine the distance to the group as a whole. If
we assume that a given target star in an external galaxy is on a
circular orbit around the center of that galaxy, three observables
--the two proper motions and the radial velocity-- suffice to
determine the three unknowns: the inclination of the orbit, the
rotational velocity and the distance. Several factors complicate the
application of this simple technique to real galaxies: 1) the target
galaxy may have substantial space-motion, 2) stars in real galaxies
are not on circular orbits (e.g. due to spiral-arm streaming motions
or induced by random motions), 3) stars in the galaxy are at
significantly different distances from the Sun (the near side of M~31
is \apx 5\% closer than the far side), 4) target stars might have
large z-heights or velocities (``run-away'' stars).
   In this paper we show how one-percent distances  can be obtaind for
the nearest spirals, even in the presence of the complications
mentioned  above.
\end{abstract}

\section{Introduction}
   At the moment, astrometry is undergoing a quiet revolution. ESA's
Hipparcos mission set the stage. Currently there are two Hipparcos++
missions in preparation: USNO leads a collaboration to build
FAME\footnote{http://aa.usno.navy.mil/fame/}, a MIDEX type mission to
be launched in 2004, and Germany is planning the
DIVA\footnote{http://www.aip.de/groups/DIVA/} mission. These
spacecraft will extend the reach of astrometry to about two kpc from
100 parsec.  Space interferometry is the next step, with the ultimate
goal to detect and characterize Earth-like planets around nearby stars
(NASA's TPF and ESA's DARWIN missions). Before that, NASA's Space
Interferometry Mission (SIM) can determine the distances to virtually
any star in the Milky~Way with an accuracy of a few percent. Many
important problems in Milky Way research can be solved with
micro-arcsecond astrometric data.

SIM will also contribute significantly to the field of cosmology.
Globular clusters play an important role in cosmology in that they
contain the oldest stars known to mankind. However, precise age
determinations are currently hampered by the lack of accurate
distances. SIM will provide such distances, and will hence establish a
firm lower limit to the age of the universe.  SIM would contribute
more directly to cosmology if its data could be used to establish the
parallax of nearby spirals and hence provide a calibration of the
zero-point of the Tully-Fisher (TF) relation. The Tully-Fisher
relation is one of the tools to measure the expansion rate (H$_0$) of
the universe. A SIM-based determination of the zero-point of the TF
relation would thus directly yield a determination of H$_0$ with an
accuracy of several percent, an order of magnitude improvement over
the current state of affairs.

Unfortunately, SIM's phenomenal precision is still not quite good
enough to achieve this. However, other, only slightly less direct
methods can be applied. In this paper we describe one of those
techniques: the  rotational parallax method.
  \section{The Rotational Parallax Method: The Circular Orbit Case}
\label{sec:The_Rotational_Parallax_Method:_The_Circular_Orbit_Case}
   Imagine a nearby spiral galaxy at distance, $D$ (in Mpc), inclined
with respect to the line of sight (by $i$ degrees) that has a rotational
speed of $V_c$ \kms.  The Andromeda galaxy (M~31, NGC~204) is such a
galaxy ($D\sim0.77, i\sim77^o$).  Its rotational speed ($V_c\sim270$)
induces a proper motion of $\mu = \frac{V_c}{\kappa D} \apx
73.9\,\muasyr$.  Here $\kappa$ is a constant that arises from the choice
of units, and $\kappa\sim4.74$ if $D$ is measured in Mpc and \ppm\ in
micro-arcsec per year.  Such proper motion is easily resolvable with an
instrument such as SIM. 

\begin{figure*}
\begin{center}
   \epsscale{.8}
   \plotone{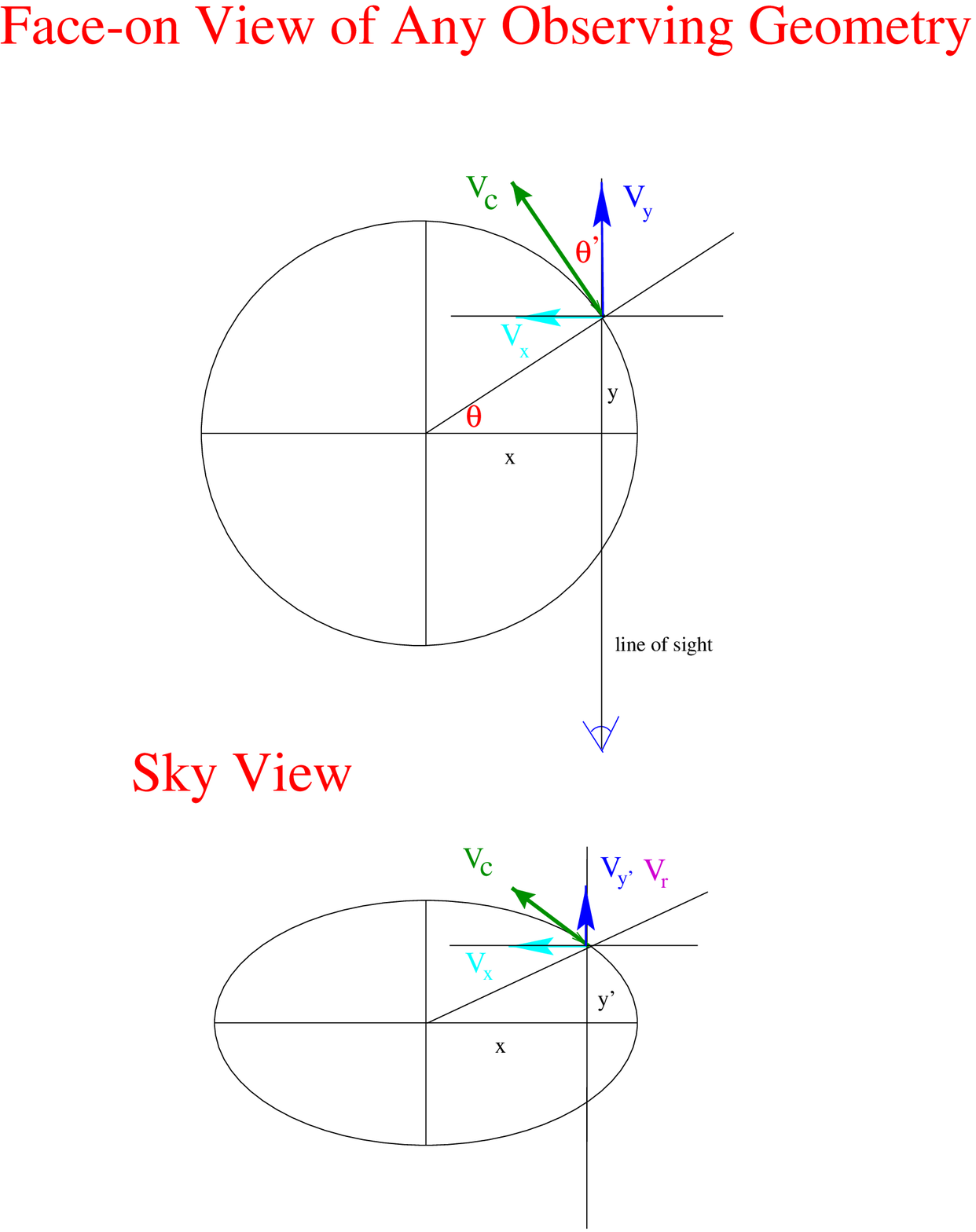}
   \figcaption{\label{fig:FO_geometry} The face-on view (top panel) of
a galaxy that is observed at an arbitrary inclination (lower panel).
The coordinate system $(x,y,y',\theta)$, the line-of-sight, as well as
the components of the circular velocity ($V_c$) are indicated.  Note
that the angles $\theta$ and $\theta'$ are identical if orbit is
circular.}
\end{center}
\end{figure*}

   In Fig.~\ref{fig:FO_geometry} we present a sketch of how a typical
nearby spiral might appear on the sky (bottom panel).  In the plane of
the galaxy (top panel) we use two coordinate systems: rectangular ($x$
and $y$) and polar ($R$ and $\theta$).  Projected on the plane of the
sky, the $x$ and $y'$ coordinates axes coincide with the major and
minor axes, respectively.  The $y'$ coordinate is the foreshortened
$y$ coordinate.

   In this section we will discuss the simplified case that all
stellar orbits are circular, more realistic situations are discussed
in \S~\ref{sec:Accurate_Distances_Q}.  The following elementary
relations between the coordinates and the various projections of the
orbital speed $\overline{V_c}$ are derived with the aid of
Fig.~\ref{fig:FO_geometry}:
\begin{eqnarray}
   x   &=& R \cos\theta \label{eqn:for_x} \\
   y   &=& R \sin\theta \\
   y'  &=& \frac{y}{\cos i} \, = \, \frac{R \sin{\theta} }{\cos i} \\
   \tan\theta &=& \frac{y}{x} \, = \, \frac{y'}{x \cos i}
      \label{eqn:tan_theta} \\
   V_x &=& -s_\Omega V_c \sin\theta' \label{eqn:V_x} \\
   V_y &=& \phantom{-} s_\Omega V_c \cos\theta' \, = \, 
      - \frac{V_x}{\tan\theta'}
      \label{eqn:V_y} \\
   V_r  &=& \phantom{-s_\Omega} V_y \sin i \label{eqn:V_r} \\
   \Vyp &=& \phantom{-s_\Omega} V_y \cos i \, = \, \kappa \, \Myp \, D
      \label{eqn:Vper}  \\
   \Vx  &=& \phantom{ -s_\Omega V_y \cos i} \, = \, \kappa \, \Mx \, D 
      \label{eqn:Vpar} \\
   \cos\theta' &=& \frac{V_y}{\sqrt{V^2_x + V^2_y}} \, = \,
                   \frac{\Myp}{\sqrt{\MypSQ + \MxSQ \cos^2 i}}
      \label{eqn:cos_theta_ell}
\end{eqnarray}
where $V_x$ and $V_y$ are the projections of $\overline{V_c}$ on the
$x$ and $y$ axes. The angle $\theta'$ between $\overline{V}_c$ and
$V_y$ equals $\theta \equiv \arctan(y/x)$ for circular orbits.  $V_r$
is the radial velocity along the line of sight and \Vx\ and \Vyp\ are
the components of $\overline{V_c}$ along the apparent major and minor
axes.  The requirement that velocities are positive in the $+x$ and
$+y$ directions leads to $s_\Omega = +1$ for counter-clockwise
rotation, and $s_\Omega = -1$ for clockwise motions.  The observable
proper motions along $x$ and $y'$ are symbolized by \Mx\ and \Myp,
respectively.
\subsection{The Principal Axes Method}
 \label{sec:The_Principal_Axes_Method}
The derivation of distances from observed proper motions and radial
velocities is a well established practice in astronomy (e.g.  orbital
parallax), and are among the most reliable distance measures
available.  Since the proper motion is the ratio of a velocity (in
\kms) and the distance, the latter can be determined from
observations.  A common problem is that the angle between the space
velocity and the line-of-sight is typically not known, so that the
distance is determined modulo $\tan{i}$.  For external galaxies, the
inclination is well established from the axis ratio of the image
and/or from the analysis of \HI\ or H~$\alpha$ radial velocity fields.

   The principal axis method is particularly appealing for the case of
circular orbits\footnote{We do not consider out-of-the plane components
here.}.  For elliptical orbits we have to introduce two additional a
priori unknown angles: the angle $\Delta \theta_M$ between the orbital
velocity ($\overline{V}_o$) and the tangent to a circular orbit at the
major axis, and the angle between a circular and an elliptical orbit at
the minor axis, $\Delta \theta_m$.  These two angular differences are
related to the eccentricity ($e$) and position angle ($\phi$) of the
elliptical orbit.  Before proceeding in the derivation of the distance,
it is profitable to re-write equations~(\ref{eqn:V_r})-(\ref{eqn:Vpar})
for stars located on the principal axes:
\begin{equation}
   \label{eqn:Eqns_mM}
\begin{array}{rcrcrcl}
V_{r,M} &=&  V_r(\cos\theta=\pm1) &=& V_y \sin i &=&
             s_\Omega V_o \cos\Delta\theta_M \sin{i}
   \\*[3mm]
\mu_{M} &=& \Myp(\cos\theta=\pm1) &=&
            \frac{\Vyp}{\kappa D} \phantom{\sin} &=&
             s_\Omega \frac{V_o \cos\Delta\theta_M \cos{i}}{\kappa D}
   \\*[3mm]
\tan\Delta_{\theta_M} &=& \frac{\Mx(\cos\theta=\pm1)}{\Myp(\cos\theta=\pm1)}
   && &&
   \\*[3mm]
\mu_m &=& \Mx(\sin(\theta)=\pm1) &=& \frac{V_x}{\kappa D} &= -&
          s_\Omega \frac{V_o \cos\Delta\theta_m}{\kappa D} \, \, ,
\end{array}
\end{equation}
where $V_{r,M}$ and $\mu_M$ are the radial and proper motion on the
major axis and $\mu_m$ equals the proper motion on the minor axis.
With an assumed inclination the first two equations of
(\ref{eqn:Eqns_mM}) can be solved for the ``major-axis'' distance
$D_M$ [eqn.~(\ref{eqn:D_M}) below].  Further simplifications can be
made in case the orbits are circular, so that the $\cos\Delta\theta_M$
and $\cos\Delta\theta_m$ terms in (\ref{eqn:Eqns_mM}) equal unity.
$D_M$ and the ``minor-major axes'' inclination and distance are given
by:
\begin{eqnarray} 
D_M &=& \frac{V_c \cos{i}}{\kappa \, \abs{\mu_M}} \, = \,
        \frac{\abs{V_{r,M}}}{\kappa \, \tan{i} \, \abs{\mu_M}}
   \label{eqn:D_M} \\
\sin{i_{mM}} &=& \sqrt{ 1 - \frac{\mu_M}{-\mu_m} }
   \label{eqn:sini_mM} \\*[3mm]
D_{mM} &=& \frac{\abs{V_{r,M}}}{\kappa \sqrt{\mu^2_m - \mu^2_M}}
   \, \, . \label{eqn:D_mM}
\end{eqnarray}
A clear advantage of the \mM\ method is that the systemic motion of
the galaxy as a whole can be taken out easily by considering points
symmetric with respect to the center of the galaxy.  Rotation-induced
proper motions at symmetric points have opposite sign, while the
radial and planar components of the systemic velocity ($V_{sys}$) have
the same orientation and magnitude.  The rotation-induced proper
motion component can thus be easily computed.  For example, when
taking out the systemic motion, $\mu_m, \mu_M$ and $V_{r,M}$ in
equations (\ref{eqn:D_M})-(\ref{eqn:D_mM}) should be replaced by
$\mu_m = \frac{1}{2} [\mu'_m(\theta=-90) - \mu'_m(\theta=+90) ]$ and
similar relations for $V_r$ and $\mu_M$. The primed relations indicate
the observed motions that include the projection of the systemic
motion.

%
%
%

One potential disadvantage of this principal axes or $nM$ method
[cf. eqn.~(\ref{eqn:D_mM})] is that the measurements of $\mu_m$ and
$\mu_M$ are to be taken at the same galactocentric radius, or from
radii with the same circular velocity and inclination.  In the
discussion above we did not consider the possibility that the actual
distances to the target stars may differ significantly, which may be
the case for the nearest galaxies (up to 5\% for M~31 \& M~33).
However, application of the major-axis method (\ref{eqn:D_M}) is
unaffected by this problem. We present a more general distance
solution in later sections.
\subsection{The Single Star Method}
 \label{sec:The_Single_Star_Method}
Another potential disadvantage of the \mM\ method is that the available
surface area in the galaxy for suitable stars is relatively small, since
the target stars are limited to regions close to the principal axes and
because suitably bright targets are quite rare.  Fortunately, other
rotational parallax varieties exist, and we discuss their merits below. 

   In fact, we can determine the distance of a single stars if we
neglect, for now, the systemic motion of the galaxy and the random
motions of the stars.  Using some of the relations
(\ref{eqn:for_x})-(\ref{eqn:Vpar}), we solve for the inclination of the
stellar orbit:
\begin{eqnarray}
   \cos{i}   &=& \frac{\Myp}{-\Mx} \tan\theta'
      \label{eqn:cosi_ellipt} \\*[3mm]
   \cos^2{i} &=& \frac{\Myp}{-\Mx} \frac{y'}{x} \, \, ,
      \label{eqn:cos2i_circ}
\end{eqnarray}
Equation~(\ref{eqn:cosi_ellipt}) holds for any type of orbit, circular
or elliptical.  Further, if the orbit is circular we can equate $\theta$
and $\theta'$ and arrive at equation~(\ref{eqn:cos2i_circ}).  The
orbital speed can be computed from the radial velocity
[eqn.~(\ref{eqn:V_r})]: $V_o = \frac{V_r}{\cos\theta'\sin{i}}$.  This
equation can be re-written for the elliptical and circular orbit cases
with the aid of eqns.~(\ref{eqn:cos_theta_ell}) and
(\ref{eqn:cos2i_circ}), respectively:
\begin{eqnarray}
V_o &=& V_r \sqrt{1 + \frac{1}{\tan^2i}} \left( 1+\frac{\MxSQ}{\MypSQ} \right)
   \label{eqn:Vo_Vr_obs} \\
V_c &=& V_r 
        \sqrt{ \frac{\Mx}{\Myp} \, \, 
               \frac{x\Myp - y'\Mx}{x\Mx + y'\Myp} }
   \label{eqn:Vc_Vr_obs}
\end{eqnarray}
In the last equation we expressed the trigonometric terms in the primary
observables. 

A rotational parallax distance can now be obtained by combining the
radial velocity and either the proper motion perpendicular or parallel
to the major axis.  As was the case for the principal axis method, we
need to make a distinction between the case of elliptical and circular
orbits.  If the orbit is elliptical, the distance must be calculated
from $V_r$ and \Myp, and depends {\em only} on the assumed inclination. 
In the circular orbit case the expressions obtained from the \Mx\ and
\Myp\ are identical and can be re-written in terms of observable
quantities only.  For the two single-star distances we find:
\begin{eqnarray}
   D_{y'} &=& \frac{V_r}{\kappa \Myp \, \tan{i}}
      \label{eqn:D_yp} \\*[3mm]
   D_{PP} &=& \frac{V_r}{\kappa} 
    \sqrt{- \frac{y'/\Myp}{x\Mx + y'\Myp}}
      \label{eqn:D_PP}
\end{eqnarray}
We will refer to distances calculated from the perpendicular and
parallel proper motions of individual stars [eqn.(\ref{eqn:D_PP})] as
``\PP\ distances.''
\subsection{Error Estimates}
 \label{sec:Error Estimates}
   It is instructive to estimate the attainable errors using SIM
astrometry.  This is most easily done for the case of circular
orbits. Below we present the expressions for the errors in the
inferred distance, inclination and circular velocity, where we express
all trigonometric terms in the primary observables:

\begin{eqnarray}
\frac{ \Delta D_{y'} }{ D_{y'} } &=&
   \sqrt{ \left(\frac{\Delta  V_r }{V_r}\right)^2 + 
          \left(\frac{\Delta \mu_{y'}}{\mu_{y'}} \right)^2 }
   \label{eqn:Delta_yp} \\*[3mm]
\frac{ \Delta D_{mM} }{ D_{mM} } &=&
   \sqrt{ \left(\frac{\Delta  V_r }{V_r}\right)^2 + 
          \left(\frac{\mu_m \Delta \mu_m}{\mu^2_m -\mu^2_M}\right)^2 + 
          \left(\frac{\mu_M \Delta \mu_M}{\mu^2_m -\mu^2_M}\right)^2   }
   \label{eqn:Delta_mM} \\*[3mm]
\frac{ \Delta \cos{^2i} }{ \cos{^2i} } &=&
   \sqrt{ \left(\frac{\Delta \Myp}{\Myp}\right)^2 + 
          \left(\frac{\Delta \Mx}{\Mx}\right)^2 }
   \label{eqn:Delta_cos2i} \\*[3mm]
\left( \frac{ \Delta V_c }{ V_c } \right)^2 &=&
   \left( \frac{\Delta  V_r }{V_r} \right)^2 + 
      \left(  \frac{y'}{2} \, 
             \frac{x (\MxSQ-\MypSQ) + 2 y' \Mx \Myp}
                  {(x \Mx + y' \Myp) (x \Myp - y' \Mx)}
      \right)^2 \times \nonumber \\*[1.5mm]
&& \phantom{  \left( \frac{\Delta  V_r }{V_r} \right)^2 + }
    \left( \left( \frac{\Delta \Mx}{\Mx} \right)^2 + 
           \left( \frac{\Delta \Myp}{\Myp} \right)^2
    \right)
   \label{eqn:Delta_Mper} \\*[3mm]
\left( \frac{ \Delta D_{PP} }{ D_{PP} } \right)^2  &=&
   \left(\frac{\Delta V_r }{V_r}\right)^2 + 
   \frac{1}{4 ( x \Mx + y' \Myp )^2 } \times \nonumber \\*[1.5mm]
&& \phantom{ \left(\frac{\Delta  V_r }{V_r}\right)^2 + } 
   \left( (x \Mx)^2 \left( \frac{\Delta \Mx}{\Mx} \right)^2 + 
          (x \Mx + 2 y' \Myp)^2  \left( \frac{\Delta \Myp}{\Myp} \right)^2
   \right)
   \label{eqn:Delta_PP}
\end{eqnarray}
At large inclination, where $x\Mx >> y'\Myp$, the errors on
$V_c$ and \DPP\ simplify to \\ $\left( \frac{\Delta V_r }{V_r}
\right)^2 + \frac{1}{4} \left( \left(\frac{\Delta
\Mx}{\Mx}\right)^2 + \left(\frac{\Delta \Myp}{\Myp}\right)^2
\right)$. 

\begin{figure*}
\begin{center}
   \plotone{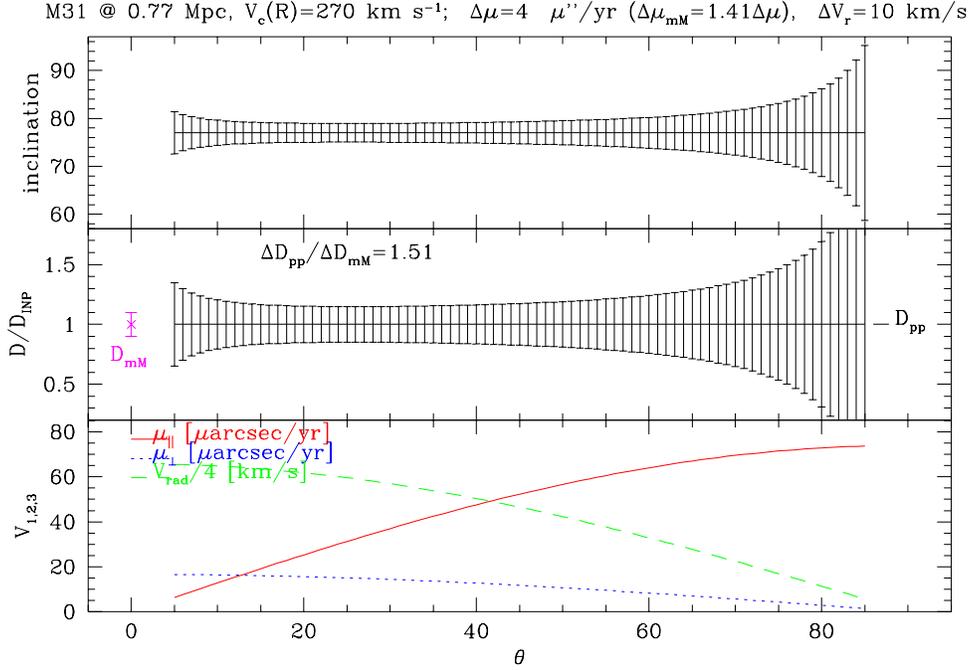}
   \figcaption{ \label{fig:M31_4mmas} Rotational parallax
considerations for M31 with 4 \muasyr\ astrometry errors. Inferred
inclination (top panel) and distances (middle panel) from the
observable proper motions and radial velocity (bottom panel). All
parameters as a function of galactic azimuth $\theta$ (=0.0 at the
major axis, =90 at the minor axis). The model input parameters are
listed above the figures. The inclination was calculated from
eqn.~(\protect\ref{eqn:cos2i_circ}). In the middle panel we plot the
principal axes distance [$D_{mM}$; eqn.~(\protect\ref{eqn:D_mM})] at
$\theta=0^o$, and the distances derived from the proper motion
parallel to the minor axis [\DPP; eqn.~(\protect\ref{eqn:D_PP})]. The
ratio of the $PP$ and $mM$ errors indicated in the middle panel
corresponds to the {\em smallest} $\Delta D_{PP}$ value, at
$\theta\sim25\arcdeg$ }
\end{center}
\end{figure*}

Typical values for proper motions may be $\mu_{x,y'} \sim 16$ \muasyr\
(40 for M31), $\Delta \mu_{x,y'} \sim 4$ \muasyr, while the radial
velocity uncertainty would be less than 10 \kms.  For M31, the
fractional errors in the radial velocity and the proper motions equal
about, 4\%, $\sim 25$\% and $\sim 10$\%, for $V_r$, \Myp, and \Mx,
respectively.  The total fractional errors on \DPP\, $V_c$ and
$D_{y'}$ are of order 15, 15, and 25\%, respectively.  These values
compare well with the exact values presented in
Fig.~\ref{fig:M31_4mmas}.

\begin{figure*}
\begin{center}
   \plotone{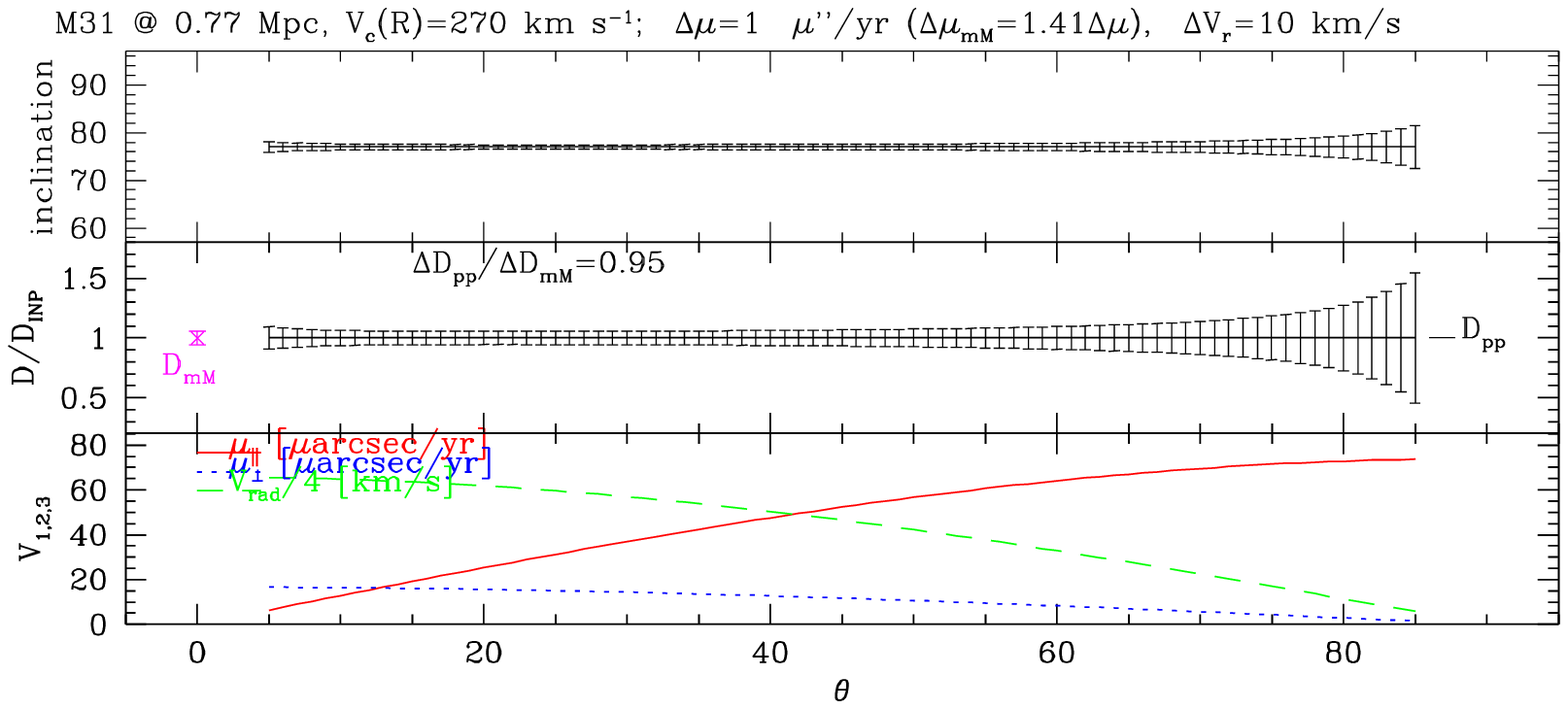}
   \vspace*{-5mm}
   \figcaption{ \label{fig:M31_1mmas} Rotational parallax
considerations for M31 with 1 \muasyr astrometry errors. Caption
as in fig.~\protect\ref{fig:M31_4mmas}. }
   \vspace*{1cm}
   \plotone{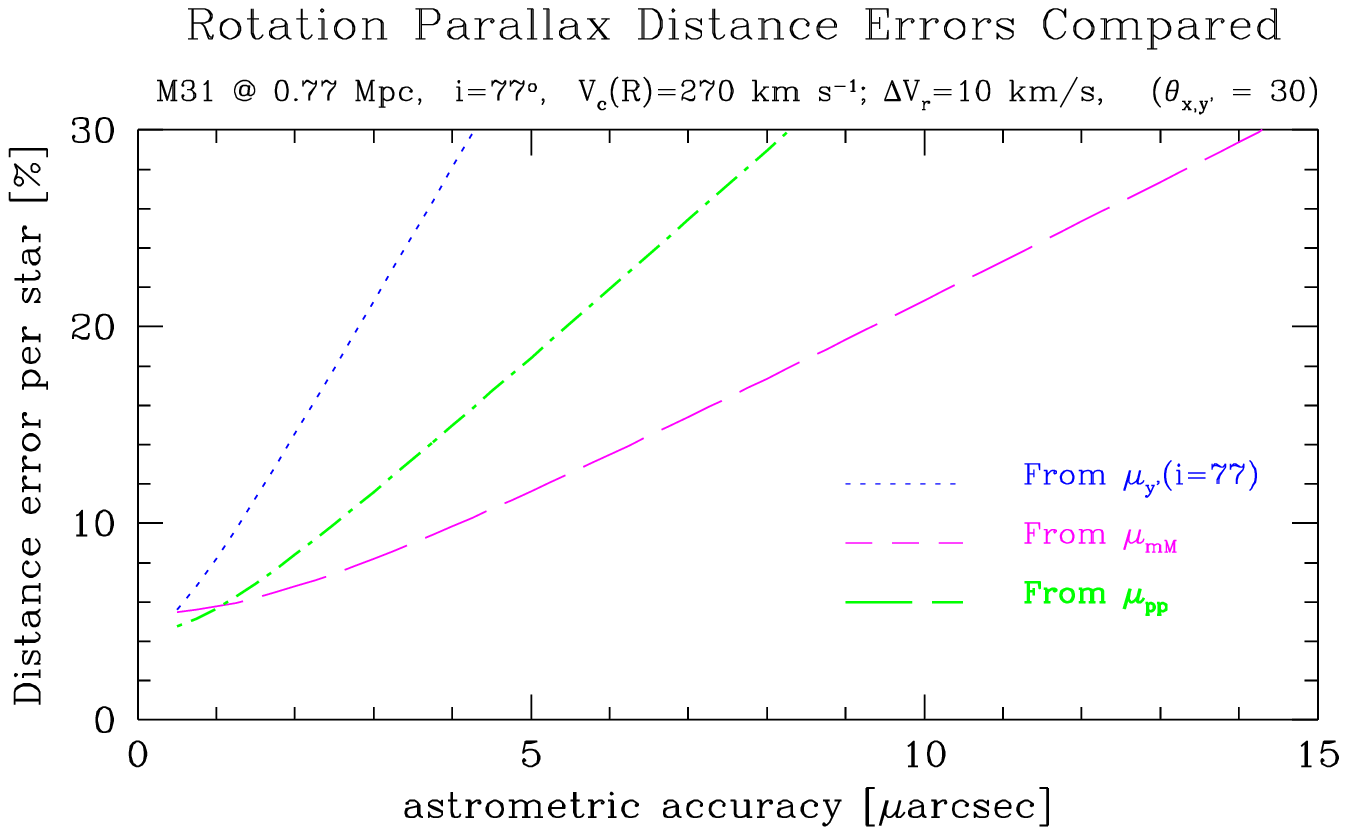}
   \figcaption{ \label{fig:M31_ast_acc} Inferred distance errors for
M31 as a function of measurement errors. For three different
estimates of $D$. The magnitude of the smallest distance error, at
$\Delta \mu = 0.5$ \protect\muasyr, is set by the assumed radial
velocity error.}
\end{center}
\end{figure*}

\begin{figure*}
\begin{center}
   \plotone{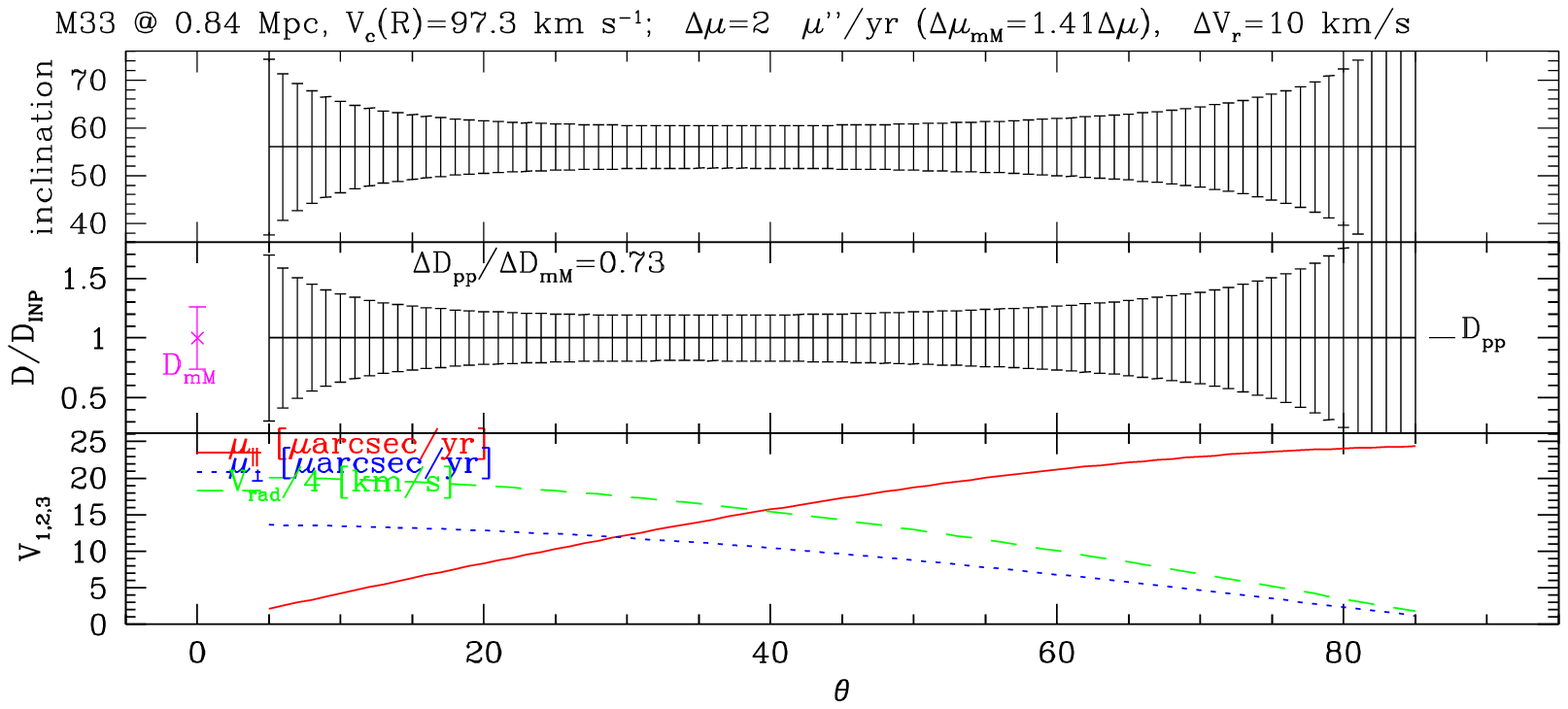}
   \figcaption{ Rotational parallax results for M33 with 2 \muasyr\
astrometry errors. Caption as in fig.~\protect\ref{fig:M31_4mmas} }
\label{fig:M33_2mmas}
   \vspace*{1cm} 
   \plotone{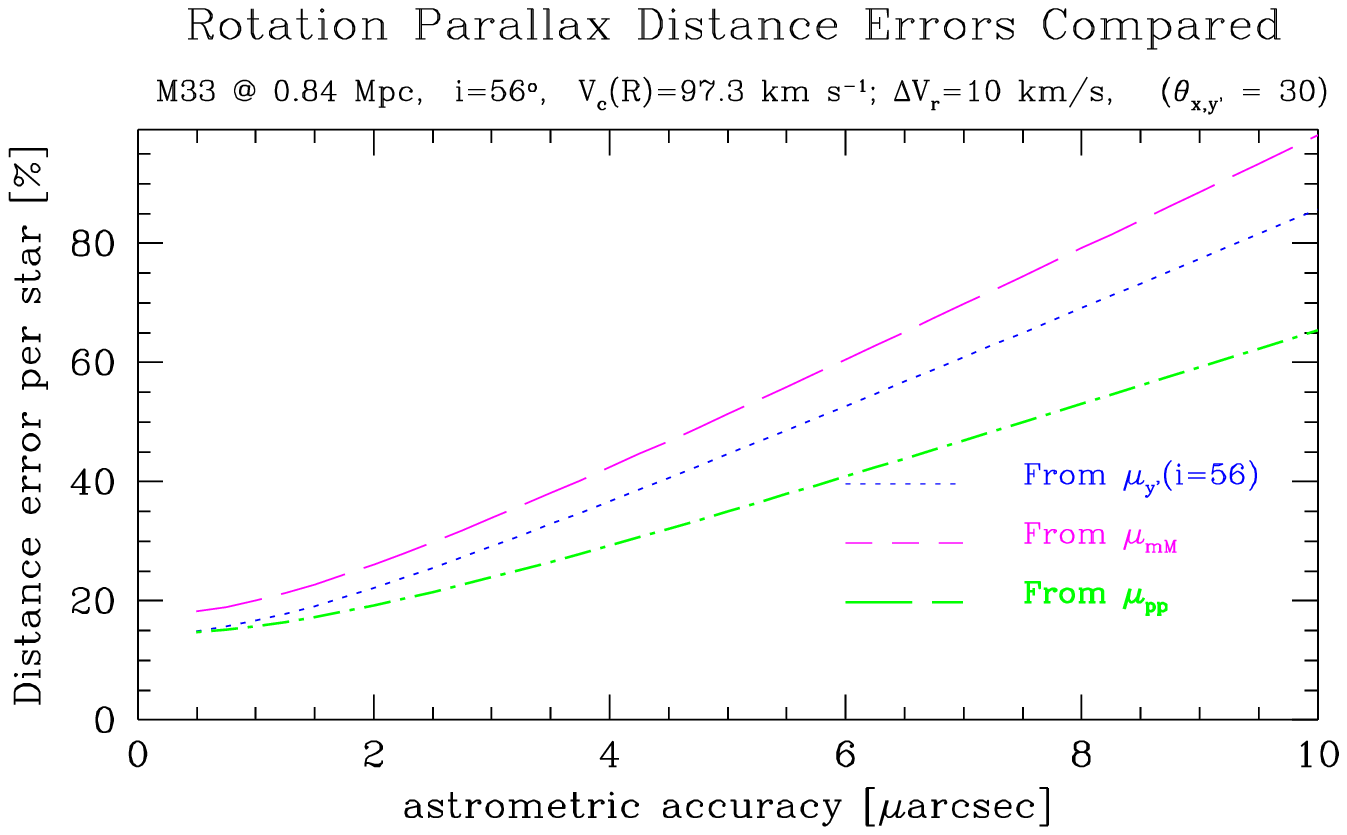}
   \figcaption{ \label{fig:M33_ast_acc} Inferred distance errors for
    M33 as a function of measurement errors. Caption as in
    fig.~\protect\ref{fig:M31_ast_acc}.}
\end{center}
\end{figure*}

\begin{figure*}
\begin{center}
   \plotone{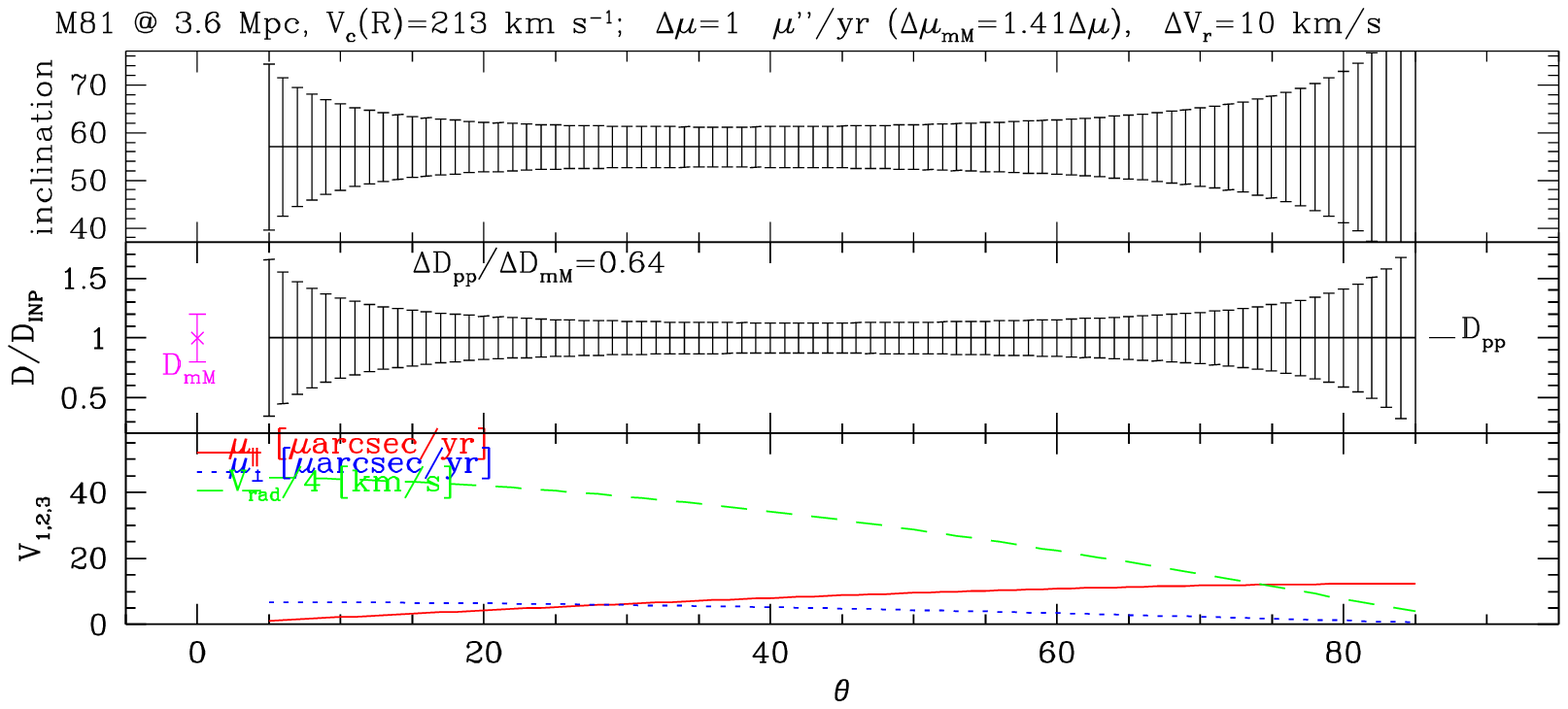}
   \figcaption{ \protect\label{fig:M81_1mmas} Rotational parallax
results for M81 with 1 \muasyr\ astrometry errors. Caption as in
fig.~\protect\ref{fig:M31_4mmas}. }
   \vspace*{1cm}
   \plotone{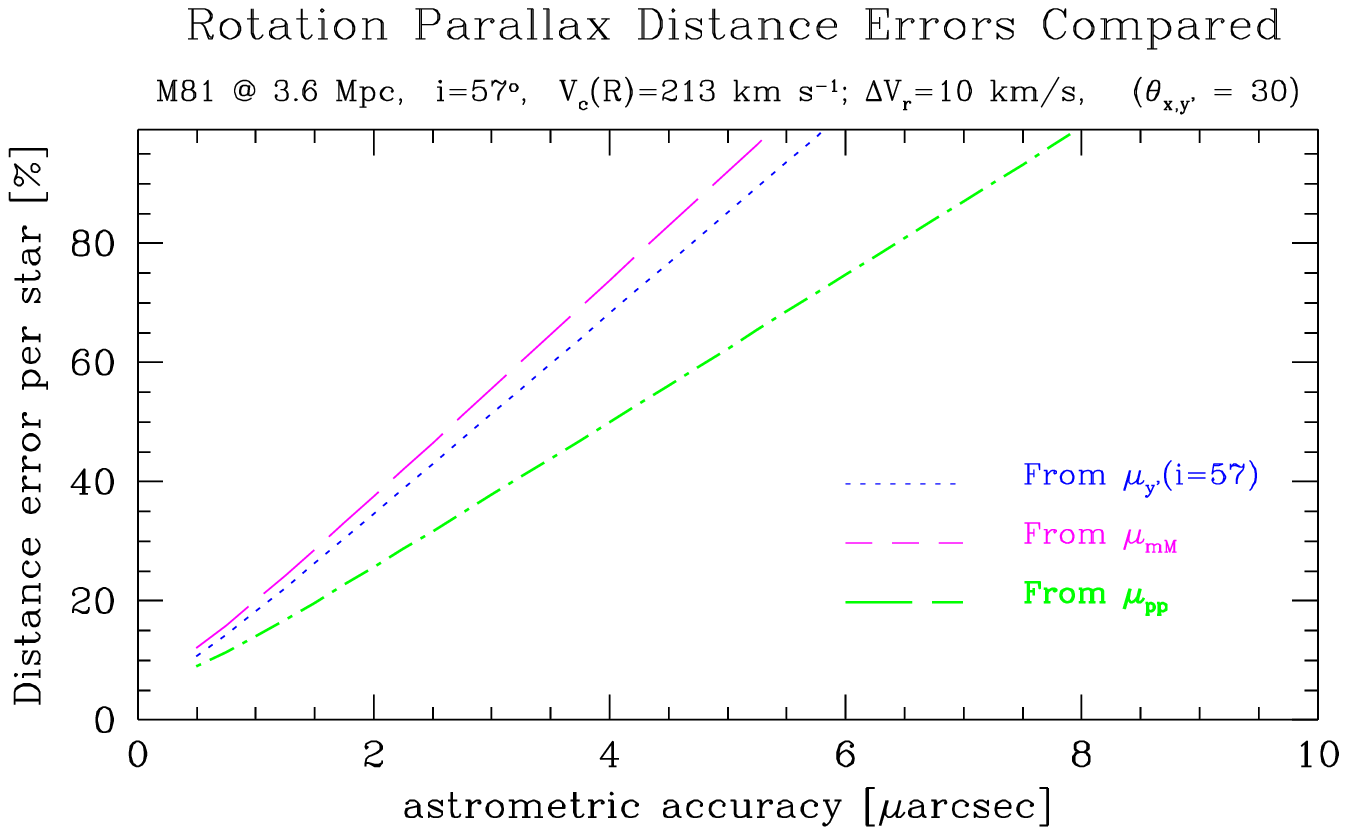}
   \figcaption{ \label{fig:M81_ast_acc} Inferred distance errors for
M81 as a function of measurement errors. Caption as in
fig.~\protect\ref{fig:M31_ast_acc}.}
\end{center}
\end{figure*}

These considerations indicate that proper motion errors dominate the
final uncertainty in the derived distances.  The \mM\ distance can have
substantially smaller errors.  Also note that, in the pure circular
motion case, the \mM\ distance estimate requires astrometry for two
stars, while the \PP\ distances can be obtained for one star
only\footnote{When comparing the final errors on the distances, we
multiply the \mM\ errors by $\sqrt{2}$ so as to compare the distance
errors per target (i.e.  per unit integration time).} .  We compare the
distance error budgets in
figures~(\ref{fig:M31_4mmas})-(\ref{fig:M81_ast_acc}), for M~31, M~33
and M~81 for a range of plausible astrometric errors. 

These figures illustrate an important aspect of the single-star
method: the attainable distance and inclination errors are not very
sensitive to the galactic azimuth $\theta$. As a result, targets from
all over the galaxy can be incorporated in the analysis. Furthermore,
because of its proximity and large rotation speed, M~31 clearly offers
the best opportunity for reliable rotational parallax measurements,
followed by M~33 and M~81.

The same figures indicate that the \PP\ method works best at low
inclinations and that the \mM\ method is superior at large
inclinations. The latter fact can be understood as follows: for a
strongly inclined galaxy, the proper motion perpendicular to the major
axis is small ($\mu_M=\mu_{y'}\sim0$), so that the $\mu_M$ term in
equation~(\ref{eqn:D_mM}) for $D_{mM}$ goes to zero. In that case
$D_{mM} \approx \abs{V_{r,M}}/(\kappa \mu_m$). As a result, the third
error-term in equation (\ref{eqn:Delta_mM}) becomes negligible. For
the \PP\ method however, both proper motions are required to derive
the distance, so that the proper motion error comes in twice.

\begin{figure*}
\begin{center}
   \plotone{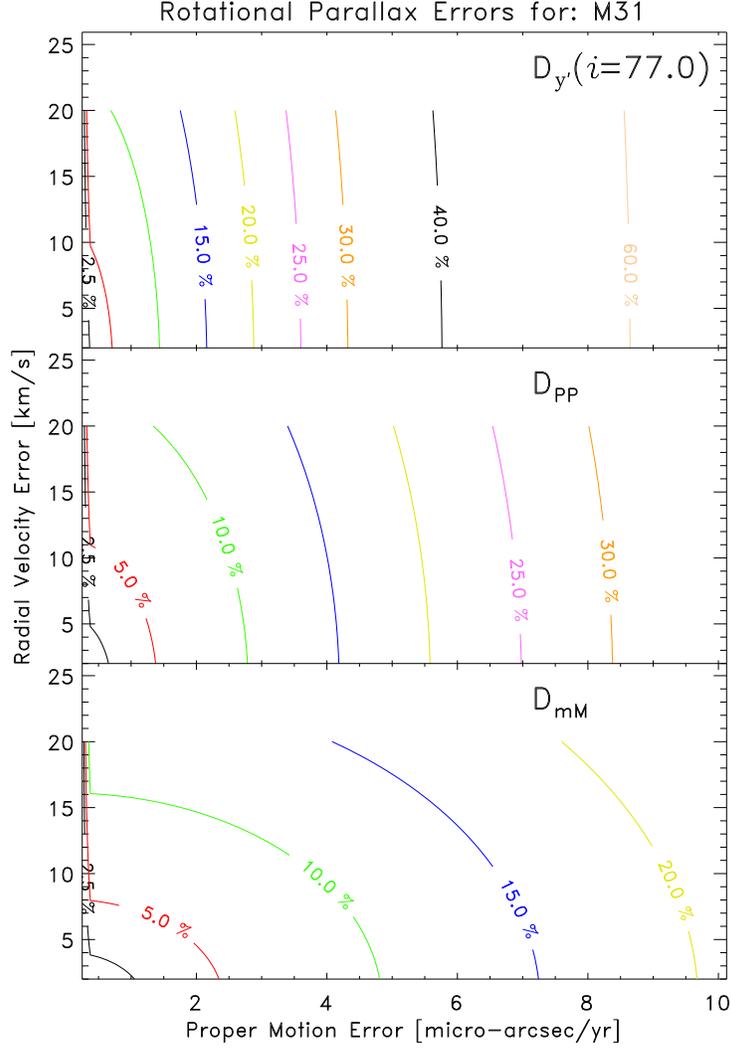}
   \figcaption{ \protect\label{fig:M31_Errcnt} Rotational parallax
errors for M~31 as a function of the astrometry errors (horizontal
scale) and the radial velocity errors (vertical scale).  In the top
panel we plot the distance as inferred for a {\em single star} from its
radial velocity and proper motion perpendicular to the galaxy's major
axis ($D_{y'}$) as given by eqn.(\ref{eqn:Delta_yp}).  In the middle and
bottom panels we plot the \PP\ and \mM\ distance errors, respectively
[cf.  eqns.~(\ref{eqn:Delta_PP}) and (\ref{eqn:Delta_mM})].  The
$D_{PP}$ and $D_{y'}$ distances were calculated for a representative
(fig.~\ref{fig:M31_4mmas}) galactocentric azimuth of $\theta=30\arcdeg$.}

\end{center}
\end{figure*}

In figure~\ref{fig:M31_Errcnt} we present the dependence of the
distance errors on both radial velocity and proper motion errors for
M~31.  In that figure we contour, from top to bottom, the fractional
errors on $D_{y'}$, $D_{PP}$ and $D_{mM}$.  For the Local Group
galaxies it will pay off to decrease the radial velocity errors to as
low values as possible. As expected, proper motion errors dominate the
error budget for distant galaxies such as M~81 ($D\sim3.6$ Mpc, not
shown).

Naively, one might expect that of our three distance estimators,
$D_{y'}$ would be determined most accurately.  After all, for this
calculation we {\em assume} an inclination and we use only {\em one}
proper potion, while $D_{mM}$ and $D_{PP}$ require an additional
proper motion measurement.  Figure~(\ref{fig:M31_Errcnt}) shows that
(for $\theta=30\arcdeg$) this is not always the case.  In fact, the
actual expected errors are complex functions of location in the
galaxy, inclination, rotation speed and distance.  For distant
galaxies, the errors are dominated by proper motion errors, so that
all three methods are equivalent (not shown).  For M~33 all methods
yield approximately equally well determined distances, while distance
errors for M~31 are minimized if we can use the principal axes method.

   For $\Delta \mu_i \sim 1$ \muasyr\ and $\Delta V_r \sim 10$ \kms, the
errors on the inferred distance equal about 6, 15, and 10\% for M~31,
M~33 and M~81, respectively.  For the Local Groups galaxies the random
motions of stars introduce additional uncertainties.  Assuming a
dispersion of 10 \kms, the effective proper motion accuracies are about
2.5 \muasyr\ at 0.8 Mpc.  The internal dispersion at the distance of
M~81 is about five times smaller and insignificant with respect to the
likely measurement errors that SIM can attain in ``reasonable''
integration times. 
  \section{Deviations from the Gedanken Galaxy}
\label{sec:Deviations_from_the_Gedanken_Galaxy}
\subsection{Systematic Motions}
 \label{sec:Systematic_Motions}
In section~\ref{sec:The_Principal_Axes_Method} we have seen that
systematic motions induced by the space motion of the galaxy
($\overline{V}_{sys}$) can be corrected for rather easily in the
principal axes method.  Likewise, a global fit to the observed proper
motions and radial velocities would allow for the determination of
\Vsysbar\ in the \PP\ method.  For example, the radial velocities of
M~31, M~33 and M~81 equal -300 \pmt 4, -200 \pmt 6 and -34 \pmt 4
\kms.  These velocities are comparable to their internal velocities of
$\sim 270$, $\sim 100$ and $\sim 213$ \kms. We will discuss this issue
in later subsections.
\subsection{The Effects of Stellar Warps}
 \label{sec:The_Effects_of_Stellar_Warps}
It is known from radial velocity observations that, for most galaxies,
the inclination gradually changes as a function of radius. This
so-called ``warping'' is most pronounced beyond the optical disk
\citep{FB90}. Although such inclination changes are hard to detect
inside the optical disk, some galaxies are known to warp in the the
outskirts of the optical disk\footnote{e.g., The Milky Way
\citep{DSL2000,PBJ1997}, and references therein, NGC~7814
\citep{LDF1995}, and others \citep{SBF1990,Fea1991}}, see
\citet{RC1998} for a recent review. These authors measured the warps
at a distance ($R_{25}$) where the B-band surface brightness is
roughly 25 magnitudes per square arcseconds. This surface brightness
is reached at radii of approximately 90 and 31 arc-minutes in M~31 and
M~33, respectively \citep{RC3}.

Cursory inspection of large-scale optical pictures of the Andromeda
galaxy suggests that the plane of the galaxy is warped in the outer
parts of the galaxy. In fact, the orientation of the outer isophotes
starts changing --an indication of a warp-- beyond about 90 arc-minutes
from the center \citep{WK1988}. 

Also, an analysis of the spiral structure and \HI\ radial velocity
field of M~31 suggests that the inclination is about 60\arcdeg\ at
10\arcmin\ from the center. Due to the almost linear warping of
$+0\arcdeg.3$ per kpc, an inclination of 80\arcdeg\ is reached at a
radius of 100\arcmin\ \citep{RB91}.  Braun's determination of the
radial inclination gradient is based on the association of \HI\ with
spiral arms, but does not self-consistently include the
non-axisymmetric velocity component that is likely to be induced by
the spiral density wave. It thus may be that the actual warp is less
severe than claimed by Braun.

Similarly, an analysis of the optical spiral arms in M~33 suggests
that the inclination of M~33 changes from forty to sixty-three
degrees across the disk \citep{SH1980}, although this result is
disputed \citep{Metal1984}.  On the other hand, the \HI\ warping of
M~33 \cite{CS1997} sets in around $R_{25}$ (i.e., 30 arcmin, or 8.2
kpc at 0.84 Mpc).
\subsection{Spiral Structure}
 \label{sec:Spiral_Structure}
Non-circular orbits, or more generally, streaming motions with respect
to the simplified picture of the circular orbit might significantly
complicate the determination of rotational parallaxes.  Streaming
motions induced by spiral density waves can reach tens of kilometers
per second, depending on the galaxy, and the locations of the target
with respect to the spiral arms.  To first order, such deviations
would induce systematic errors in the inferred distances that equal
the systematic velocities induced by spiral density waves, or about
$\Delta V/V_c \sim20$\%.

The theory of spiral arm density waves has been developed extensively
[e.g.  \citep{LYS69}].  Applications of this theory to radial velocity
fields of external galaxies show that streaming motions of the order of
20-50 \kms\ [e.g.  \citep{V80,VKS88,BV92,TA93}].  The analysis of the
spiral structure of M~31 by \citet{RB91} indicates streaming motions of
40 \kms\ for radii smaller than 40$\arcmin$, of 20 \kms\ between
30$\arcmin$ and 75$\arcmin$ and of order 10 \kms\ in the outermost
parts.  M~31 exhibits a well developed two-armed spiral pattern with a
small, but radially varying pitch angle \cite{RB91}.  In the inner
region, $R\la 27\arcmin$, the spiral has a larger pitch angle ($\phi
\sim 16\arcdeg$) than in the outer region ($\phi \sim 7\arcdeg$). 

Stars will respond to the potential induced by the spiral density
wave.  If the pattern is logarithmic, the radial and tangential
velocities [$\tilde{V}_R(R)$ and $\tilde{V}_\theta(R)$] induced by the
perturbation can be found analytically.  These perturbation velocities
vary with (extra)~galactocentric radius $R$.  So do the pitch angle
and the spiral phase ($\chi$).  $\tilde{V}_R$, $\tilde{V}_\theta$,
$\phi$ and $\chi$ can be calculated from the rotation curve, the
stellar velocity dispersion, and the pattern-speed and amplitude of
the perturbation, where the last three properties are not well
established observationally for our target galaxies.  This procedure
has been followed for the Milky Way
\citep{CM73,AL97,MZp97,MZ99,LMD2000}, and can be readily generalized
to external galaxies.

It is also possible that the spiral structure is generated in response
to a bar or nearby companions.  For the Andromeda galaxy, the
companions, M~32 and NGC~205, have estimated peri-centers between 13 and
35 kpc [e.g.  \citep{gB767778,SS86,CB88}].  In fact, radial velocities
and SIM-based proper motions can conceivably be used to {\em determine}
the parameters of the spiral pattern, and allow for detailed tests of
various theories of spiral structure. 

However, these spiral-structure theories have a large number of
parameters.  In the density-wave theory, almost all factors that
determine the spiral pattern depend on galactocentric radius, so that
these ``factors'' are really {\em functions} with many more unknowns.
Further, on small scales, the details of spiral structure may deviate
from the large-scale $n$-armed spiral.  For example, it has been
suggested that the large star-formation complex NGC~206 in Andromeda
resulted from the recent interaction between two spiral-arm segments,
moving with relative velocity of about 30 \kms\ \citep{Mea97}.
Alternatively, the often seen bifurcations in the spiral patterns of
spiral galaxies may arise from the superposition of spiral modes with
different multiplicity.  The Milky Way may be an example of
two-plus-four armed spiral galaxy \citep{AL97}.

Clearly, we would like to avoid the complications that arise due to
streaming motions on small and large scales.  We will discuss the
effects of non-axisymmetric streaming motions in more detail in
section~\ref{sec:Practical_Implementation}.
  \section{Accurate Distances, Notwithstanding Perturbations?}
\label{sec:Accurate_Distances_Q}
In the previous sections we discussed the case of circular orbits and
alluded occasionally to the case of elliptical orbits.  A complication
of elliptical orbits is that the angle between the orbital velocity
and the line-of-sight ($\theta'$) can not be deduced from the position
of the target and the inclination of the galaxy.  That is to say,
$\theta' \ne \theta$ [c.f. eqns.~(\ref{eqn:tan_theta}) and
(\ref{eqn:cos_theta_ell})].  This means that predicting the orbital
motion (velocity and direction) at point ($x_2,y_2$) given measurements
at ($x_1,y_1$) becomes significantly more complicated, even if these
points lie at the same galactocentric radius.  Knowledge of the
ellipticity and position angle of the orbit are required to solve this
problem.  Furthermore, due to the intrinsic dispersions of the target
population, additional velocity vectors are added to the motions of
the targets.  All these parameters are likely to depend on
galactocentric distance.  And finally, the systemic motion of the
galaxy needs to be determined and its effects subtracted from on the
individual radial velocities and proper motions.
\subsection{\Myp-\Vr\ Correlations}
  \label{sec:Myp-Vr_Correlations}
Here we extend on the procedure to determine the $D_{y'}$ distance
[\S\S~\ref{sec:The_Single_Star_Method}, eqn.~(\ref{eqn:D_yp})] and
make use of the fact that the radial velocity and the $V_{y'}$
velocity are two orthogonal components of the {\em total} space
velocity in the $r-y'$ plane ($V_{tot,ry'}$, see
figure~\ref{fig:SO_Vels}).  Like in equation~(\ref{eqn:D_yp}), the
distance then follows from the observed radial velocity and $y'$
proper motion: $D_{y'} = V_r / (\kappa \Myp \tan i_t)$, where $i_t$ is
the angle between $V_{tot,ry'}$ and the plane of the sky.  Generally
speaking, $i_t$ differs from the geometric inclination of the galaxy.
For example, $i_t$ will lie between $i$ and 90\arcdeg\ if random
motions are unimportant and the systemic motion is due to Hubble flow
only.  We illustrate the various contributions to $V_{tot,ry'}$ in
figure~\ref{fig:SO_Vels}.  From this figure we deduce the following
relations for the observed $y'$ proper motion and the radial velocity:
\begin{eqnarray}
V_{y'}    &=& 
   \left( V_{o,y} \phantom{\, + V_{e,y}} + V_{\sigma,y} \right) \cos i 
   + V_{sys,ry'} \cos i_s + V_{\sigma,z} \sin i
   \label{eqn:Vyp_r} \\*[5mm]
V_{obs,r} &=&
   \left( V_{c,y} + V_{e,y} + V_{\sigma,y} \right) \sin i 
   + V_{sys,ry'} \sin i_s - V_{\sigma,z} \cos i 
   \label{eqn:Vobs_r}
\end{eqnarray}
where $i_s$ represents the angle between the systemic motion and the
plane of the sky.  The orbital velocity in the $y$-direction ($V_{o,y}$)
comprises both a circular orbit term ($V_{c,y}$) and a contribution for
the ellipticity of the orbit ($V_{e,y}$).  The random motions in the $y$
and $z$ directions are denoted as $V_{\sigma,y}$ and $V_{\sigma,z}$,
respectively.  We now proceed by expressing the observable $y'$ proper
motion in terms of the observable radial velocity.  To this end we solve
eqn.~(\ref{eqn:Vobs_r}) for $(V_{o,y}+V_{\sigma,y})$, substitute the
result in (\ref{eqn:Vyp_r}) and divide by $\kappa D$ to arrive at:
\begin{eqnarray}
\mu_{y'} &=& \frac{V_{obs,r}}{\kappa D \tan i} +
   \frac{1}{\kappa D} 
    \left(
       V_{sys,ry'} ( \cos i_s - \frac{\sin i_s}{\tan i} ) +
       \frac{V_{\sigma,z}}{\sin i}
    \right) \label{eqn:mu_yp_Vr} \\*[5mm]
  &=& \Ayp V_{obs,r} + \beta_{y'}
  \, \, . \label{eqn:mu_yp_AB} 
\end{eqnarray} 
These two equations show that the observable $y'$ proper motion is a
linear function of the observable radial velocity.  We recognize the
$\cos i_s$ and $\sin i_s$ terms in (\ref{eqn:mu_yp_Vr}) as the $y$ and
radial velocity components of the systemic velocity, respectively.
More importantly, equation~(\ref{eqn:mu_yp_Vr}) shows that the slope
$\Ayp$ is {\em independent} of the presence of non-circular motions.
This is so because in eqns.~(\ref{eqn:Vobs_r})-(\ref{eqn:mu_yp_AB}) we
use the total $y$ component of the orbital speed: for the
purpose of distance determination, it is not necessary to know how the
total $y$ velocity is distributed between the circular, elliptical and
random components.
\subsection{\Mx\ Correlations}
  \label{sec:Mx_Correlations}
   Following the method of the previous subsection, we will express
the observed \Mx\ as a function of either the observed radial velocity
or the $y'$ proper motion.  In order to do so, we need to connect the
$x$ and $y$ components of the orbital speed via a simple
relation. This is straightforward for the circular component of the
orbits. As above, the total observable velocity comprises circular
velocities ($V_{c,x}$ and $V_{c,y}$), ``elliptical velocities''
($V_{e,x}$ and $V_{e,y}$), random motion components and the systemic
velocity.  We also employ the $r$ and $y'$ motions derived above
[eqns.~(\ref{eqn:Vyp_r}) and (\ref{eqn:Vobs_r})] and the $x$ motion:
\begin{eqnarray}
V_{tot,x} &=& V_{c,x} + V_{e,x} + V_{sys,x} + V_{\sigma,x}
   \label{eqn:Vtot_x}
\end{eqnarray}
First we solve eqn.~(\ref{eqn:Vobs_r}) for $V_{c,x}$, substitute the
result in (\ref{eqn:Vtot_x}) and divide by $\kappa D$ to find \Mx:
\begin{eqnarray}
\mu_{x,r} &=& 
   - \frac{\tan\theta}{\sin i \kappa D} \Delta V_{obs,r} + 
   \frac{\tan\theta}{\kappa D} 
   \left( V_{e,y}+V_{\sigma,y} - \frac{V_{\sigma,z}}{\tan i} \right) + 
   \frac{V_{e,x} + V_{sys,x} + V_{\sigma,x} }{\kappa D}
   \, \, , \label{eqn:mu_tot_xr}
\end{eqnarray}
where $\Delta V_{obs,r} \equiv V_{obs,r} - V_{sys,ry'} \sin i_s$.
Likewise, when solving the $y'$ proper motion for $V_{c,x}$ and using
$\Delta \Myp \equiv \Myp - V_{sys,ry'} \cos i_s$, we find:
\begin{eqnarray}
\mu_{x,y'} &=& - \frac{\tan\theta}{\cos i}\Delta \Myp + \frac{\tan\theta}{\kappa D} 
   \left( V_{e,y}  + V_{\sigma,y} + V_{\sigma,z} \tan i  
   \right) +
   \frac{V_{e,x} + V_{sys,x} + V_{\sigma,x} }{\kappa D} \, \, .
   \label{eqn:mu_tot_xyp}
\end{eqnarray}
In case the orbits are exactly circular,
equations~(\ref{eqn:mu_tot_xr}) and (\ref{eqn:mu_tot_xyp}) simplify
to:
\begin{eqnarray}
\mu_{x,r} &=& \Mx  = 
    -\frac{y'/x \Delta V_{obs,r}}{\sin i \cos i \kappa D}
   + \frac{V_{sys,x} }{\kappa D}
   \, = \,  \Axr (y'/x \Delta V_{obs,r}) + \beta_x
   \, \, ,
   \label{eqn:mu_tot_x_Vr} \\*[5mm]
\mu_{x,y'} &=& \Mx = 
    -\frac{ y'/x \Delta \Myp }{\cos^2 i} \phantom{xii}
   + \frac{V_{sys,x} }{\kappa D}
   \, = \,  \Axyp (y'/x \Delta \Myp) + \beta_{x,y'}
   \, \, ,
   \label{eqn:mu_tot_x_Myp} 
\end{eqnarray}
where we have replaced $\tan\theta$ by $y'/(x\cos i)$. The independent
variables $(y'/x \Delta V_{obs,r})$ and $(y'/x \Delta \Myp)$ are
defined for $x=0$ since, for circular orbits, they are proportional to
$(V_c \sin\theta)$ [cf.
eqns.~(\ref{eqn:tan_theta})-(\ref{eqn:Vpar})].  However, the random and
elliptical contributions to $\Delta V_{obs,r}$ can ``blow-up'' close
to the minor axis as a result of the multiplication by $1/x$. It is thus
advisable to down-weight the regions close to the minor axis. Like the
$\Ayp$ slope derived in the previous subsection, the slope $\Axr$
depends on both the inclination and the distance, but in an
independent manner. The correlation between \Mx\ and \Myp\ yields the
inclination of the galaxy directly.

The proper motion equations (\ref{eqn:mu_yp_AB}),
(\ref{eqn:mu_tot_x_Vr}), and (\ref{eqn:mu_tot_x_Myp}) represent the
multi-star equivalent of the single-stars case and include the systemic
motion of the galaxy: we solve them for inclination and distance in \S
\ref{sec:The_Rotational_Parallax_Distance} below, and we discuss the
effects of non-circular and random motions in
section~\ref{sec:Practical_Implementation}.
\subsection{The Rotational Parallax Distance}
 \label{sec:The_Rotational_Parallax_Distance}
In the previous subsections we found that the $y'$ and $x$ proper
motions can be expressed in terms of the observed radial velocities.
From these linear slopes the inclination and distance follow:
\begin{eqnarray}
\Ayp &=& \frac{1}{\kappa D \tan i}
   \label{eqn:alpha_yp} \\*[3mm]
\Axr  &=& \frac{-1}{\kappa D \tan i \cos^2 i }
   \label{eqn:alpha_x}  \\*[3mm]
\cos^2 i &=& \frac{\Ayp}{-\Axr} \, = \, \frac{1}{-\Axyp}
   \label{eqn:cos2i_alphas} \\*[3mm]
D_{\alpha i,xy'}&=& \frac{1}{\kappa \Ayp \tan i} \, \approx \,
                   \frac{-1}{\kappa \alpha_{xr} \tan i \cos^2 i }
   \label{eqn:D_alphas_inc} \\*[3mm]
D_{\alpha xy'r} &=& 
   \frac{1}{\kappa \Ayp} \sqrt{\frac{-\Ayp}{\Axr + \Ayp}}
   \, \, . \label{eqn:D_alphas}
\end{eqnarray}
The above relations hold exactly for circular orbits.  We will
investigate the effects of non-circular orbits in the next subsection. 
Equations (\ref{eqn:cos2i_alphas}) and (\ref{eqn:D_alphas_inc}) for $i$
and $D_{\alpha i, xy'}$ are the equivalent of the relations for
inclination and distance as derived in the single-star method
[(\ref{eqn:cos2i_circ}) and (\ref{eqn:D_yp})].  It thus follows that
$\Axr$ is equivalent to $x\Mx$, $\Ayp$ corresponds to $y'\Myp$ and
$\Ayp \Delta V_{obs,r} \leftrightarrow \Myp$, so that
eqn.~(\ref{eqn:D_alphas}) is obtained. 
\subsection{The Systemic Velocity Vector}
 \label{eqn:The_Systemic_Velocity_Vector}
A straightforward manner to determine the systemic velocity of the
galaxy would be to take the appropriate averages of the observed proper
motions and radial velocities of targets that have similar $|x|$ and
$|y|$ coordinates. In this way, the motions induced by the internal
motions of the galaxy are averaged out, so that the components of the
systemic velocity of galaxy become apparent. However, due to
elliptical streaming motions these values can only be considered to be
``reasonable'' values. A better determination of the systemic velocity
is possible from a full solution of equations~(\ref{eqn:mu_yp_Vr}),
(\ref{eqn:mu_tot_xr}) and (\ref{eqn:mu_tot_xyp}).
  \section{Practical Implementation}
\label{sec:Practical_Implementation}
In section~\ref{sec:Deviations_from_the_Gedanken_Galaxy} we have
reviewed the various effects that complicate the implementation of the
single-star rotational parallax method. Above we outlined  the
modifications required to measure the systemic motion. In the next
subsections we show that neither warping of the stellar disk nor
spiral structure inhibits our ability to determine accurate rotational
parallaxes.
\subsection{Inclination Effetcs}
 \label{sec:Inclination_Effetcs}
For many spiral galaxies, it is possible to determine the inclination
from the axis-ratio of contours of constant optical surface
brightness. For M~31 this procedure is more complicated due to its
large inclination and finite thickness.  However, the inclination can
also be determined from \HI\ velocity fields [e.g., see \citet{U1983}
and \citet{BB1984} for M~31 and \citet{CS1997} for M~33]. These
inclination estimates typically have an uncertainty of a few degrees.
If such \HI\ inclination estimates were to be used to determine the
distance from the slope of the observed \Myp-\Vr\ relation, the
resulting distance uncertainties would be of order 20\% for M~81 and
M~33, and a factor of two for M~31. To obtain distance estimates that
are accurate at the percent level, the other proper motion component
needs to be utilized.

  From equations~(\ref{eqn:mu_yp_AB}), (\ref{eqn:mu_tot_x_Vr}), and
(\ref{eqn:mu_tot_x_Myp}) it is clear that no a-priori knowledge of the
inclination is required, {\em if} the inclination does not vary
significantly with radius. In that case, distance and inclination can
be determined, even when the targets are arbitrarily located across
the face of the galaxy, provided that the kinematic variations are
sufficiently sampled.  On the other hand, if the inclination does
change with radius, it might be best to select targets in an
elliptical annulus with an axis-ratio equal to the cosine of the
best-guess inclination. The drawback of such an approach is that the
range in radii sampled will increase if the a-priori inclination
estimate was wrong\footnote{By about a factor of two for a 2\arcdeg\
error at $i_{est}=75\arcdeg$.}. In principle, the best distance
determination is possible when all targets have similar distances from
the galaxy center so as to minimize the effects of any possible radial
variations of inclination (and rotation speed).

\subsection{Non-Circular Motions}
 \label{sec:Non_Circular_Motions}

   In general, non-circular motions will be present in our target
galaxies. The question is, how will those motions affect our ability
to determine an accurate rotational parallax.
Equations~(\ref{eqn:mu_tot_xr})-(\ref{eqn:mu_tot_x_Myp}) above show
that elliptical motions in both the $x$ and $y$ direction can
adversely influence the distance determination. Below we will outline
a technique that can be used to detect those motions and correct for
their effects.

The contribution from elliptical motions in
equations~(\ref{eqn:Vobs_r}) and (\ref{eqn:Vtot_x}) can be written as
the sum of two component. If the angular variation of an elliptical
streaming component is identical to the azimuthal dependence of the
circular velocity component we term that component ``invisible.'' That
is to say, an invisible elliptical streaming field will induce a
proper motion and radial velocity field that is indistinguishable from
that of the circular streaming field.  Thus, the invisible elliptical
motion is given by:
\begin{eqnarray}
\overline{V}_{ei} &=& V_{ei,x} \sin\theta \, \hat{x} + 
            V_{ei,y} \cos\theta \, \hat{y} \, ,
   \label{eqn:Ve_invisible}
\end{eqnarray}
with $\hat{x}$ and $\hat{y}$ the unit vectors in the $x$ and $y$
directions, respectively.  Also note that, if $V_{ei,x}$ equals
$V_{ei,y}$, then the resulting elliptical motion amounts to an
additional circular velocity term, and should be absorbed in
$V_c$. All elliptical streaming components orthogonal to
$\overline{V}_{ei}$ are directly detectable in the observed stellar
motions. The invisible elliptical streaming can significantly bias the
inferred inclination and distances since their proper motion and
radial velocity signature are indistinguishable from the circular
motion terms. In section~\ref{sec:The_Invisible_Components} we show
how invisible streaming motions can be detected, and their
distance-bias corrected for.

The exact functional form of the elliptical streaming field depends on
the physical mechanism that drives such non-circular motions (\S
\ref{sec:Spiral_Structure}). A full investigation of the dependence of
the effect of non-circular streaming on the accuracy with which the
rotational parallax can be determined is beyond the scope of the
current work. However, we will explore the effects of non-circular
motion by investigating the effects of a toy model for elliptical
streaming. This illustrative model has the invisible streaming
component discussed above plus an additional orthogonal, visible,
component:
\begin{eqnarray}
\overline{V}_{e,toy} &=& \Veib + V_{ev,x} \cos\theta \, \hat{x} + 
            V_{ev,y} \sin\theta \, \hat{y} \, ,
   \label{eqn:Ve_toy}
\end{eqnarray}
with $V_{ev,x}$ and $V_{ev,y}$ the visible $x$ and $y$ components of
the elliptical streaming field, respectively.
\subsection{The Works}
 \label{sec:The_Works}
The rotational parallax distances derived above can be improved upon
if the \Myp-\Vr\ and \Mx-\Vr\ equations [(\ref{eqn:mu_yp_Vr}) and
(\ref{eqn:mu_tot_x_Vr})] are solved simultaneously. In order to arrive
at a unbiased solutions in case \Veib\ is non-zero, additional
constraints are required. For example, one could demand that the
inferred radial gradients of the inclination and rotation curve are
small and linear. However, better, non-parametric additional
constraints are available: 1) the lack of azimuthal variation of the
rotation curve inferred from \Myp, \Vr\ and \Mx, and 2) the
requirement that the rotation speed derived from these three
observable are identical, to within the errors\footnote{Since \Vr\ is
easier to measure than \Myp\ (for M~31), and the resulting
$V_c(\theta)$ should be equivalent, we will only use the information
contained in \Vr\ and \Mx.}

Such a multiple non-linear regression solution as described above is
beyond the scope of the current paper. However, we will investigate a
poor-man's approach to the problem and use that to derive estimates of
the accuracy to which the rotational parallax can be determined with an
instrument like SIM.  Although the multiple regression approach is
clearly advisable to make optimal use of the available data, the
poor-man's approach has the advantage that it clearly illustrates the
problematic areas of the rotational parallax determination
method. Further, since the poor-man's route will not lead to the best
possible solution, the error estimates so obtained are likely to be
improved upon when using a multiple non-linear regression technique.

In the poor-man's approach we split the procedure in four distinct
steps. In the first step, the \Myp-\Vr\ correlation
[eqn.~(\ref{eqn:mu_yp_Vr})] is used to determine a solution for the
product of distance and the the tangent of the inclination:
$D\times\tan i = (\kappa \alpha_{y'})\rtp{-1}$
[cf. eqn.~(\ref{eqn:D_alphas_inc})]. As discussed in section \S
\ref{sec:Myp-Vr_Correlations}, this determination of $D\times \tan i$
has no sensitivity to elliptical streaming motions {\em at all}. In
the second step we determine $D, i$ and $\overline{V}_{ev}$ from the
\Mx-\Vr\ correlation [eqn.~(\ref{eqn:mu_tot_x_Vr})] given the
previously determined value for $D\times\tan i$. In the third step, we
repeat the previous step $N_{try}$ times. We randomly selected a value
for $D \times \tan i$, based on its estimated value and dispersion
obtained in step \#1. The final best value for the to-be-fitted
parameters is obtained by averaging the $N_{try}$ results, where the
derived errors equal the second moment of the $N_{try}$ values. In the
fourth and final step, we check the additional constraints that the
inferred rotation curve should have negligible azimuthal dependence,
and that the rotation speed value inferred from the radial velocities
and proper motions are equal to within the errors (see the Diagnostics
section below for details).  In principle, the diagnostics step can
already be incorporated in step \#2 to improve the estimates on the
parameters. We will not do so in the spirit of deriving upper limits
to the error budgets.

\subsubsection{Rotation Speed Diagnostics}
    \label{sec:Rotation_Speed_Diagnostics}

Starting from equations~(\ref{eqn:Vtot_x}) and (\ref{eqn:Vobs_r}) we
write\footnote{We set the contributions from the systemic motion and
the stellar random velocities to zero.}:
\begin{eqnarray}
V^{inf}_{c,x}(\theta) &\equiv&
   \left( V_{c,x} + V_{ei,x} \right) \, \sin\theta
   \, = \,   -\left[   \kappa D \mu_{obs,x}  - V_{ev,x}(\theta) \right]
   \label{eqn:Vc_inf_x} \\
V^{inf}_{c,r}(\theta) &\equiv&
   \left( V_{c,r} + V_{ei,y} \right) \, \cos\theta
   \, = \,  
   \phantom{-\kappa D [} \frac{V_{obs,r}}{\sin{i}} - V_{ev,y}(\theta)
   \, \, , \label{eqn:Vc_inf_r}
\end{eqnarray}
where $V^{inf}_{c,x}(\theta)$ and $V^{inf}_{c,r}(\theta)$ are the
circular velocities inferred from the observed $x$ proper motions and
the radial velocities, respectively. These equations clearly
illustrate that the invisible component of the elliptical streaming
motions are indistinguishable from the circular velocity term.  Here
we retained the general expressions for the visible components of the
elliptical streaming field. The right-hand sides (RHSs) of these two
equations can be constructed from the observed \Vr\ and \Mx\ motions
and the fitted values for distance, inclination and \Vevb. If the
correct values for $i$ and $D$ are used, and if the true visible
streaming field is subtracted from the RHSs,
equation~(\ref{eqn:Vc_inf_x}) should show a pure sine modulation,
while only the cosine modulation should contain significant power in
equation~(\ref{eqn:Vc_inf_r}).

In general, neither the true inclination and distance, nor the correct
elliptical streaming field will be used in the empirical determination
of the RHSs of the $V^{inf}_c$ equations. Thus, higher order $\theta$
modulations will be observable in the RHSs of the $V^{inf}_c$
equations.  The usage of an erroneous inclination ($i_e$) is most
damaging because the estimated position angle, $\theta_e$, is
determined from the estimated inclination such that $\theta_e =
\arctan(\frac{y'}{x \cos{i_e}})$. In practice, a search for additional
modulations in the $V^{inf}_c$ equations will be performed in terms of
the estimated azimuth, not the true azimuth ($\theta_t$). To see what
the consequences are, we suppose that the true inclination, $i_t$, can
be expanded to first order around $i_e$. With $\Delta i \equiv
i_e-i_t$ we find:
\begin{eqnarray}
\sin\theta_t &\approx& 
   \left( 1 -  \frac{\Delta i}{4} \tan i_e \right) \sin\theta_e -
               \frac{\Delta i}{4} \tan i_e  \sin{3\theta_e}
   \label{eqn:sin_thetha_true} \\
\cos\theta_t &\approx& 
   \left( 1 +  \frac{\Delta i}{4} \tan i_e \right) \cos\theta_e -
               \frac{\Delta i}{4} \tan i_e  \cos{3\theta_e}
   \label{eqn:cos_thetha_true}
\end{eqnarray}
At the inclination of the Andromeda galaxy, the $3\theta_e$
modulations have an amplitude of about $\frac{2\Delta
i}{1\,\rm{degree}}$ percent of the $\theta_e$ amplitudes, or several
\kms. We find that, with a sufficient number targets, such effects are
easily detectable.

In fact, if we expand the RHSs of the $V^{inf}_c$ equations in
Fourier series, their $3\theta_e$ coefficients can be used to calculate
new estimates for the inclination:
$
\Delta i_c = A_{c3} \frac{4}{\tan i_e}
$ and 
$
\Delta i_s = A_{s3} \frac{4}{\tan i_e}
$
with $A_{c3}$ and $A_{s3}$ the measured Fourier coefficients of the
$\cos 3\theta_e$, and $\sin 3\theta_e$ modulations. Significant
$3\theta_e$ coefficients can occur in two cases: 1) the correct
elliptical streaming was determined but the estimated inclination is
wrong, and 2) the right inclination was determined, but the wrong
$3\theta_e$ components was fitted for the elliptical streaming field.
The observed $3\theta_e$ terms can only be used to arrive at a better
inclination estimate in case the correct \Vevb\ has been
subtracted. Generally speaking, that can not known to be the
case. However, in both cases, the presence of significant $3\theta_e$
terms indicate that the model used is inadequate and that a better
solution is possible. Further, the ambiguity of the meaning of any
detected $3\theta_e$ modulation also illustrates the necessity of a
multiple non-linear regression technique.

\subsection{The Invisible Components}
 \label{sec:The_Invisible_Components}

So far, the invisible component remains undetectable, even when the
$3\theta_e$ constraints are included, and even if a multiple
non-linear regression technique has been used.  As it turns out, the
lowest order ($\theta_e$) Fourier coefficients of the expansion of the
$V^{inf}_c$ equations provide powerful additional constraints on the
invisible elliptical streaming components.

For example, suppose that \Veib\ is non-existent, in that case, the
$A_{c\theta}$ and $A_{s\theta}$ coefficients of the cosine and sine
modulations should be identical and equal to the rotation speed of the
galaxy. If \Veib\ is non-zero, $A_{c\theta}$ will not be equal to
$A_{s\theta}$. Unfortunately, equations~(\ref{eqn:Vc_inf_x}) and
(\ref{eqn:Vc_inf_r}) are not sufficient to determine the three
unknowns contained in the two equations. However, it is not necessary
to know the values of $V_{ei,x}$ and $V_{ei,y}$ separately. Only the
algebraic sum of the two invisible components, $V_{ei,xy}(\theta)
\equiv (V_{ei,x} \sin\theta + V_{ei,y} \cos\theta)$, occur in
equations~(\ref{eqn:mu_tot_xr}) and (\ref{eqn:mu_tot_xyp}). Thus, in
these two equations, $V_{ei,xy}(\theta)$ can be replaced by
$(A_{s\theta} + A_{c\theta})-V_c(\sin\theta + \cos\theta)$. In a
multiple regression technique, the circular velocity thus becomes a
to-be-fitted parameter. As a result, the $V^{inf}_c$ equations will
also allow for the determination of both invisible components of the
elliptical streaming field.

The arguments presented above show that, with the additional
constraint that the $V^{inf}_c$ equations only show a $\theta$
modulations, all parameters of the model can be determined. Because
even the ``invisible'' components of the elliptical streaming field
can be determined experimentally, the inferred distance, inclination
rotation speed and galaxian space motion can be measured without
significant systematic errors.

\subsection{Final Accuracies}
 \label{sec:Final_Accuracies}

In the SIM book it is suggested one-hundred targets are observed in
both M~31 and M~33 and twenty-five in M~81. Our discussion above
indicate that systematic errors will be unimportant so that the final
attainable distance error is inversely proportional to the square-root
of the number of target stars.  With such a moderate number of targets
and a proper motion uncertainty of 4\muasyr, extremely accurate
distances can be determined for the nearest spirals: $\Delta D \sim$
0.7\%, 1.9\% and 10\% $(\times \frac{\Delta \mu}{4 \muasyr})$, for
M~31, M~33 and M~81, respectively.

It should be kept in mind that these considerations neglect several
important aspects that will be subject of future study. First,
systematic effects {\em internal} to the galaxies such as spiral arm
streaming motions or runaway stars will reduce the final achievable
results. On the positive side, it will be possible to identify
``deviant'' objects using the single star method so that they can be
eliminated from the target list. Further, it may very well be possible
that a global fit to the proper motions and radial velocities of the
targets will produce significantly tighter results if we impose
smoothness criteria for the radial variation of rotation speed and
inclination. It might also be possible to use external information on
the radial gradients of the inclination and rotation curve from the
\HI\ velocity field to further decrease the errors. This too will be
investigated in the near future.

We have generated numerical models that describe the stellar motions
in three potential SIM targets: M~31, M~81 and M~3. The stellar disks
of these model galaxies are inclined by 77\arcdeg, 56\arcdeg, and
57\arcdeg\ with respect to the line-of-sight, respectively. The
motions of the stars have circular velocity components of 270, 213 and
97.3 \kms, while we add a random component to the targets of 10 \kms\
in all three directions. We computed several models with either fixed
inclinations or a small inclination gradient. We also varied the
elliptical streaming component, from non-existent to strong
(approximately 20\% of the circular velocity), where we tried several
angular dependencies consistent with the toy model described in
section~\ref{sec:Non_Circular_Motions} (i.e., $\theta$ components
only). Our method of analysis of this model data is described by the
poor-man's approach of section~\ref{sec:The_Works}. We ran six
different models with 100-600 targets. Each of these models were
``observed'' $N_{try}=200$ times where for each try we added random
terms to the stellar space motions to simulate observational
errors. To the radial velocity we added random errors from 2.5 to 10
\kms, in steps of 2.5 \kms, while we used a large range of proper
motion errors (between 1 and 50 \muasyr).

\begin{figure*}
\begin{center}
   \centerline{
      \epsfxsize=15.0cm \epsfysize=18cm
      \epsfbox{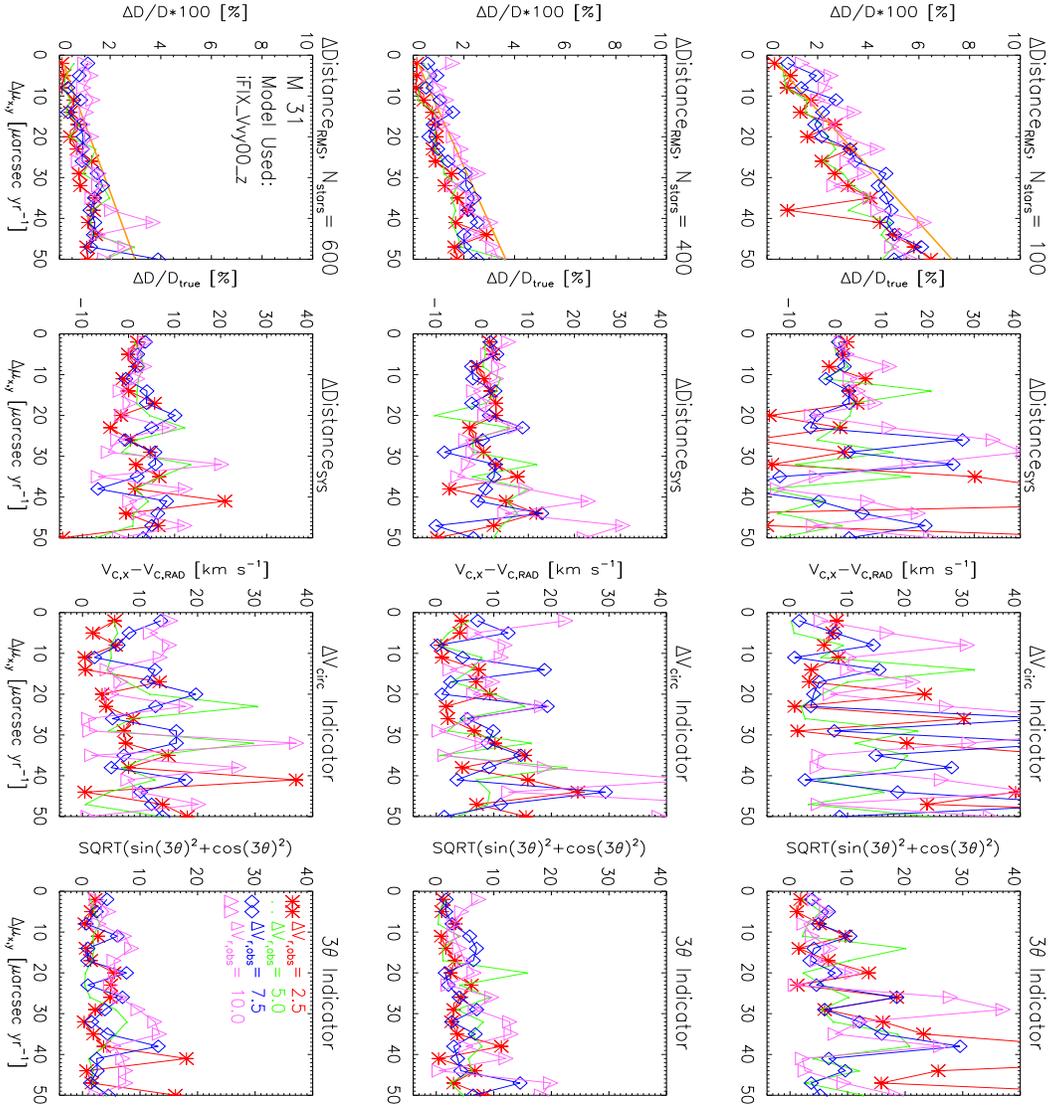}
   }
   \figcaption{\label{fig:M31_Mod_Res} The results of ``measurements''
of simulated M~31 models (see \S \ref{sec:Final_Accuracies} for
details). The results for 100, 400 and 600 targets are displayed in
the top, middle and bottom rows. In the left two columns we plot the
rms and systematic distance errors, respectively. The two systemic
error indicators are plotted in the two right columns. In each panel,
four cases are plotted with observational errors between 2.5 and 10
\kms. In the ``RMS'' error plot, we also draw a line (thick, orange)
derived from the \mM\ error formula [eqn.~(\ref{eqn:Delta_mM})] with
$N_{stars}$ targets.}
\end{center}
\end{figure*}

\begin{figure*}
\begin{center}
   \centerline{
      \epsfxsize=15.0cm \epsfysize=18cm
      \epsfbox{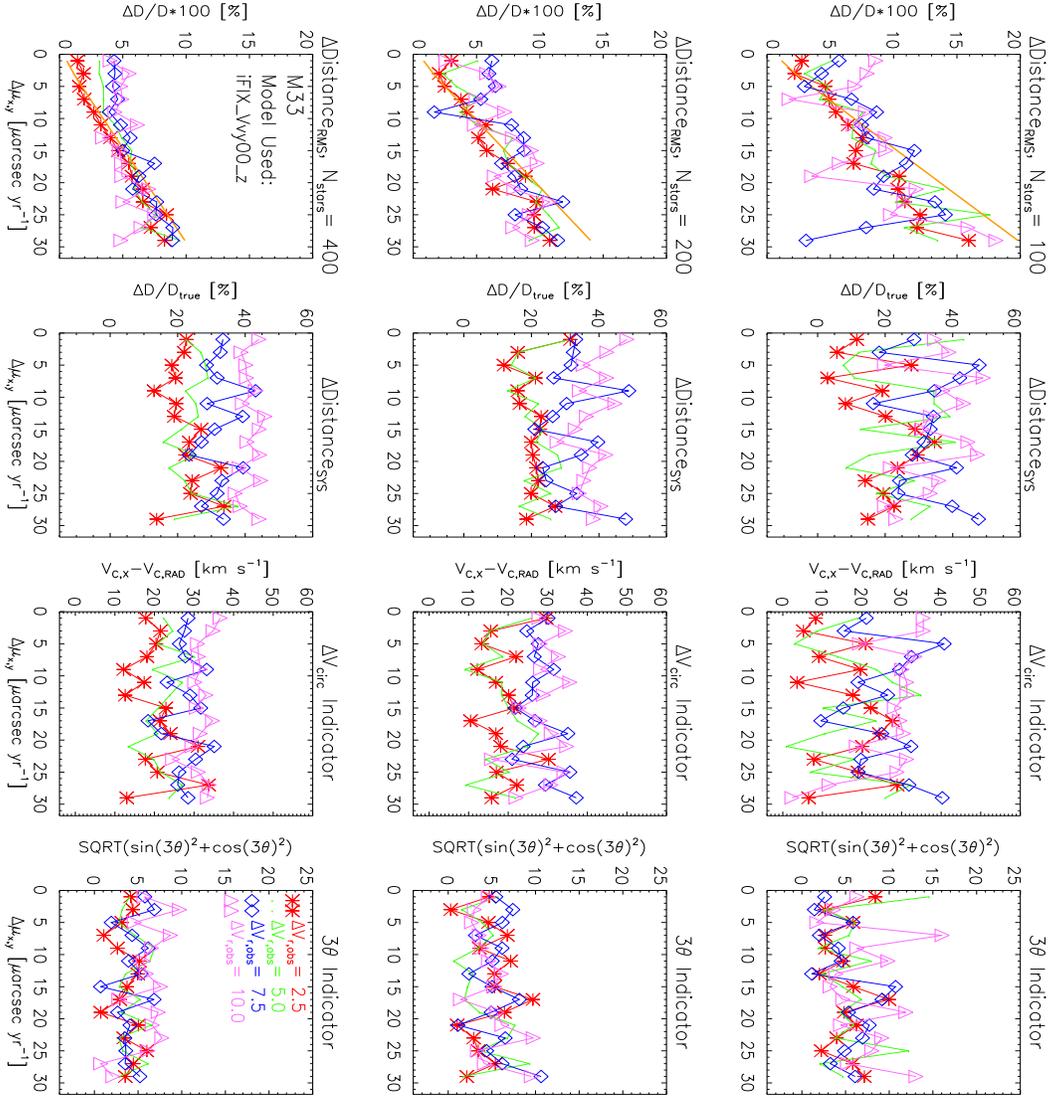}
   }
   \figcaption{\label{fig:M33_Mod_Res} Same as for
figure~\ref{fig:M31_Mod_Res}, but for M~33. The number of simulated
targets equals, 100, 200 and 400, from top to bottom.}
\end{center}
\end{figure*}

\begin{figure*}
\begin{center}
   \centerline{
      \epsfxsize=15.0cm \epsfysize=18cm
      \epsfbox{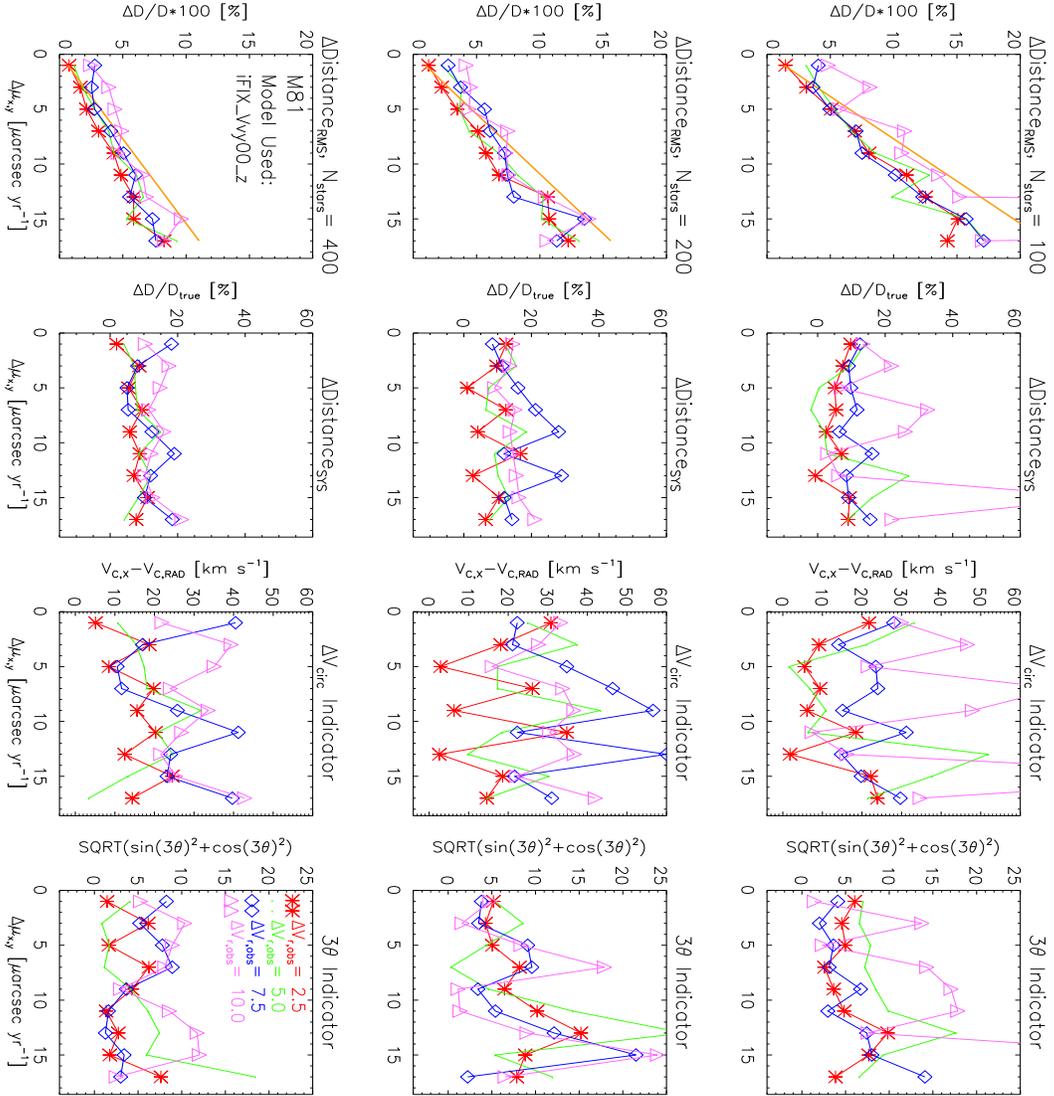}
   }
   \figcaption{\label{fig:M81_Mod_Res} Same as for
figure~\ref{fig:M31_Mod_Res}, but for M~81. The number of targets is
as for M~33.}
\end{center}
\end{figure*}

The results for the three galaxies are summarized in
figures~\ref{fig:M31_Mod_Res}-\ref{fig:M81_Mod_Res}. Each of these
three figures contain three rows and four columns. The results for the
smallest number of targets are displayed in the top row, those for the
largest number of stars in the bottom panels. In the left two columns
we plot the inferred rms and systematic distance errors,
respectively. The two systemic error flags (SEFs) are plotted in the
two right columns. For M~31, for which the observed proper motions are
largest, the rms errors are encouragingly small, even for a small
number of targets. The inferred systematic errors are also small,
which is confirmed by the small SEF values. For the other two
galaxies, significant systematic errors result from the poor-man's
approach. This is probably a result of the fact that these galaxies
have smaller rotational proper motions. On the positive side, the SEFs
are also clearly raised, indicating that the poor-man's solutions are
erroneous.

\subsection{Beyond SIM}
 \label{sec:Beyond_SIM} 

If the systemic velocity is zero, the systematic variation of of the
distancew $D$\footnote{$D(R,\theta)=\sqrt{[D(R=0) + R\sin\theta \sin
i]^2 + [R\cos\theta]^2}$} along an annulus will introduce a deviation
from the linear behavior of eqn.~(\ref{eqn:mu_yp_AB}).  However, the
deviation due to the ``proximity effect'', $\Delta \Myp$, is rather small:
$\Delta \Myp \sim4, \sim0.14, \sim0.09$ and $\sim0.01$ \masyr\ for the
LMC, M~31, M~33 and M~81, respectively.  The constant $\beta_{y'}$ in
eqn.~(\ref{eqn:mu_yp_AB}) has a distance dependence as well so that
the systemic velocity terms will also contribute to $\Delta \Myp$.  In
fact, because the contribution to $\Delta \Myp$ from internal motions
is down-weighted by the $1/\tan i$ term, the systemic contribution to
$\Delta \Myp$ tends to outweigh the former.  However, for
``reasonable'' values of the systemic velocity, $\Delta \Myp$ will be
still be rather small.

With the possible exception of the Magellanic Clouds, we do not expect
that it will be possible to determine the proximity effect with SIM
data, in part due to the small number of stars that SIM can measure.
Recall that a velocity dispersion of 10 \kms\ corresponds to about 3
\masyr, or about twenty times the proximity effect, at the distance of
M~31.  Assuming that measuring  $\Delta \Myp$ at ten \Vr\ positions
would suffice to determine the proximity effect, we estimate that at
least $10 \times 20^2 = 4000$ targets are required.  This would take
about 6,000 hours or about 20\% of the science time available during
the SIM mission.


\clearpage

\clearpage

\begin{figure*}
\begin{center}
   \epsscale{.8}
   \plotone{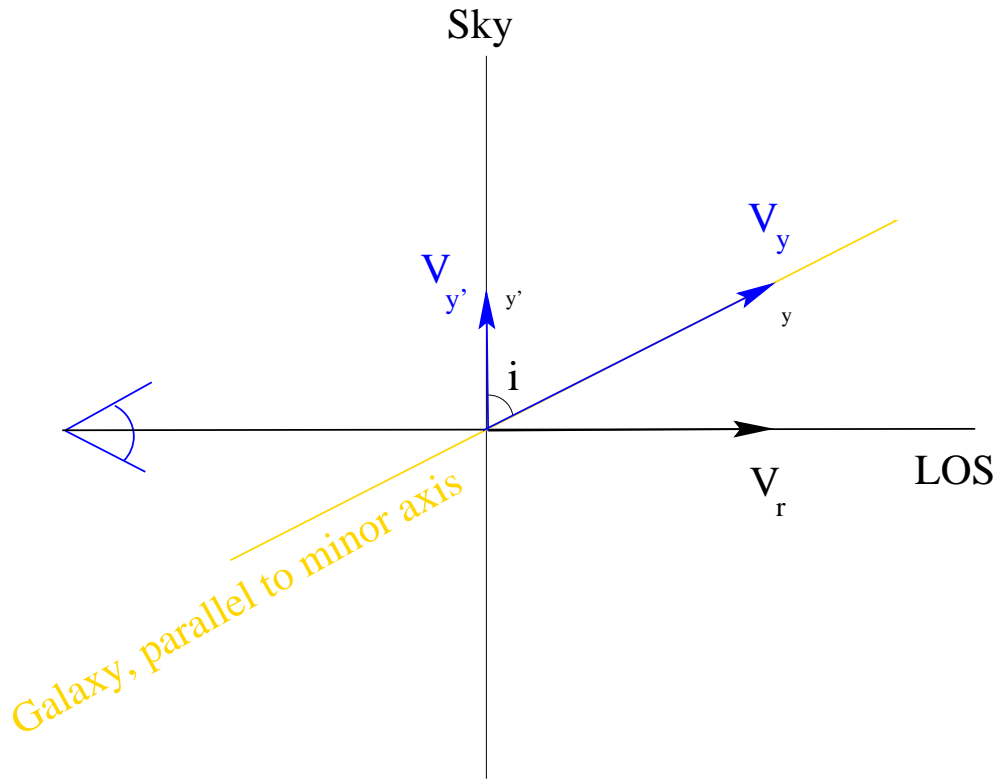}
   \figcaption{\label{fig:SO_geometry} The side-on view of a galaxy
that is observed at intermediate inclination. The x-axis points out of
the paper, the z-axis lies in the $r-y'$ plane, perpendicular to the
$y$ axis.}
\end{center}
\end{figure*}

\clearpage

\clearpage

\clearpage

\clearpage

\clearpage

%
%
%
%
%

\clearpage

\begin{figure*}
\begin{center}
   \epsscale{.8}
   \plotone{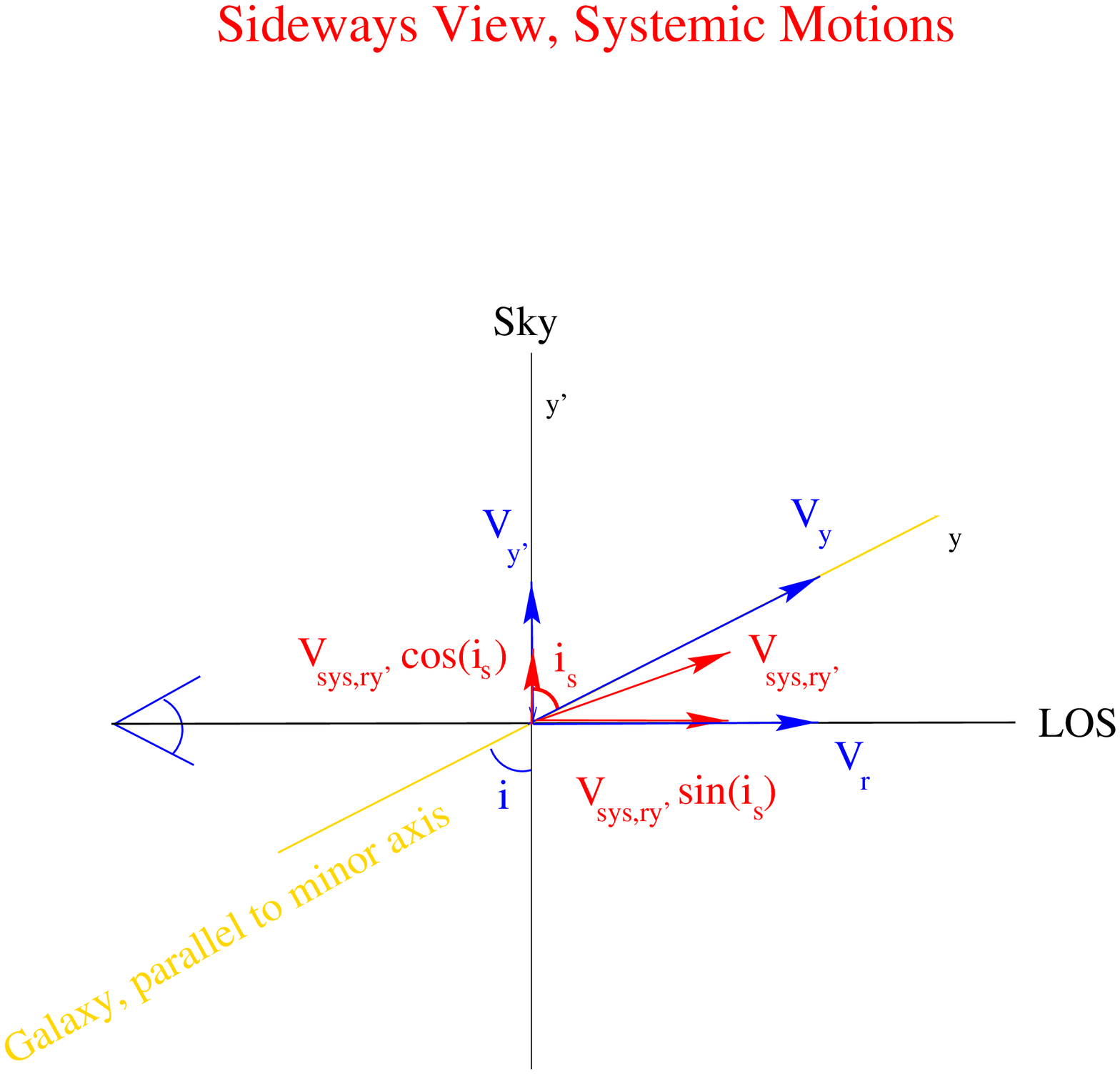}
   \figcaption{\label{fig:SO_Vels} The side-on view of a galaxy that
is observed at intermediate inclination.  In addition to the
components resulting from circular motions as in
fig.~(\ref{fig:SO_geometry}) we include (the observable projections
of) three other components: 1) the systemic velocity in the $r-y'$
plane ($V_{sys,ry'}$, red), 2) a contribution from the ellipticity of
the orbit ($V_{e,y}$), 3) a random motion term in the plane of the
galaxy ($V_{\sigma,y}$) and 4) the random motion component out of the
plane ($V_{\sigma,z}$).  The systemic motion $V_{sys,ry'}$ is inclined
by $i_s$ degrees with respect to the plane of the sky.  The total
$ry'$ velocity equals the vector sum of orbital, systematic and random
motion terms, and is inclined by $i_t$ degrees with respect to the
plane of the sky.  }
\end{center}
\end{figure*}

\clearpage

\begin{figure*}
\begin{center}
   \centerline{
      \epsfxsize=8.0cm \epsfysize=15cm
      \epsfbox{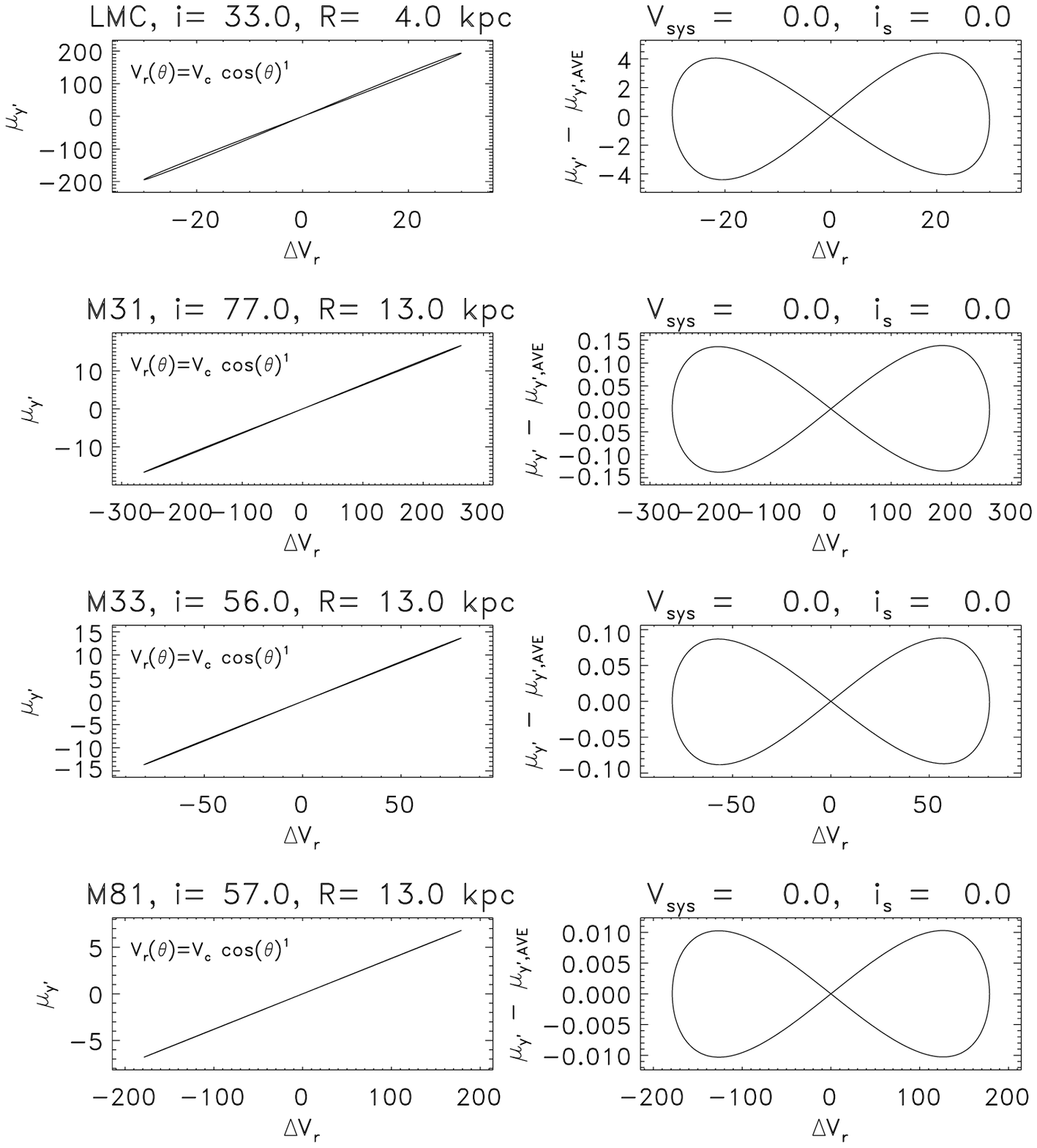}
      \hspace*{5mm}
      \epsfxsize=8.0cm \epsfysize=15cm
      \epsfbox{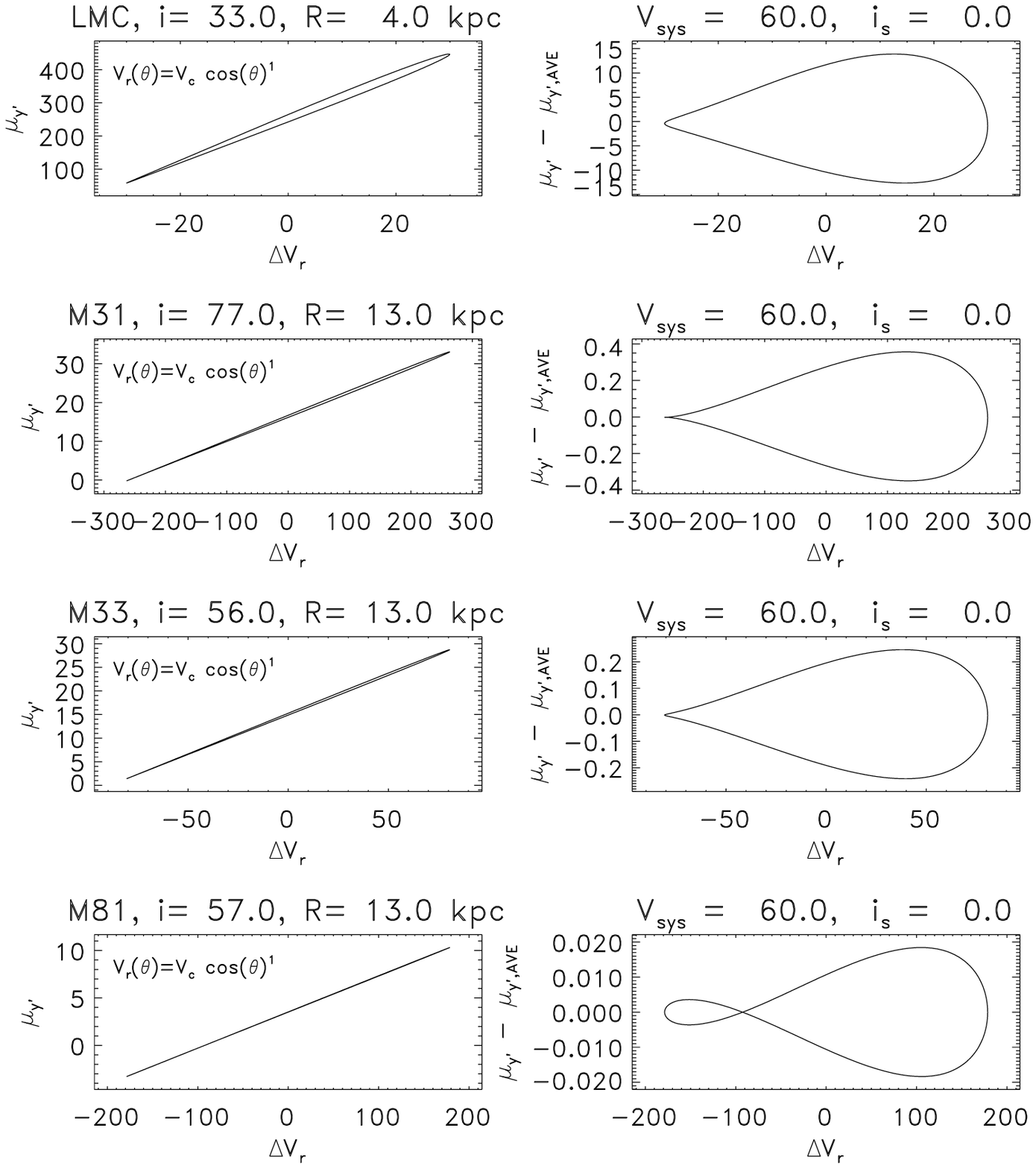}
   }
   \figcaption{\label{fig:Gals_Dvar} The proper motion $\mu_{y'}$ as a
function of $\Delta V_r \, [\equiv V_r-V_{r,sys}$] calculated
according to eqn.~(\ref{eqn:mu_yp_Vr}). The orbital velocities are
based on circular motion along a ring of radius 13 kpc: $V_r = V_c
\cos\theta$. We present two sets of panels: in the left hand set the
systemic motion of the galaxies is set to zero, the right hand set was
calculated for $V_{sys,y'}=60$\kms. In each panel, we plot two
columns. In the left column we plot $\mu_{y'}$ as a function of
$\Delta V_r$, in the right figures we have subtracted the average
regression.  The four rows are, from top to bottom, the LMC
($V_c=55$), M~31 ($V_c=270$), M~33 ($V_c=97.3$), M~81 ($V_c=213$)
simulated at the inclination indicated.  As predicted by
eqn.(\ref{eqn:mu_yp_Vr}), the regression line is straight and $D
\times \tan i$ can be determined. The deviation from the regression
line allows, in principle, to determine the distance to the
galaxy. However, the amplitude will be too small for a reliable SIM
determination, except possibly for the LMC.}
\end{center}
\end{figure*}

\clearpage

\clearpage

\clearpage

\clearpage


\begin{thebibliography}{}
%
\bibitem[Amaral \& Lepine (1997)]{AL97} 
   Amaral L.H. \&  Lepine J.R.D.,
   1997, \mnras, 286, 885 
%
\bibitem[Boulanger \& Viallefond (1992)]{BV92} 
   Boulanger F. \& Viallefond F., 
   1992, \aap, 266, 37 
%
\bibitem[Braun (1991)]{RB91} 
   Braun R., 
   1991, \apj, 372, 54
%
\bibitem[Brinks \& Burton (1984)]{BB1984} 
   Brinks E. \& Burton W.B.,
   1984, \aap, 141, 195
%
\bibitem[Briggs (1990)]{FB90} 
   Briggs F.H., 1990, \apj, 352, 15 
%
\bibitem[Byrd (1978)]{gB78}  Byrd G.G., 1978, \apj, 226, 70
\bibitem[Byrd (1977)]{gB77}  Byrd G.G., 1977, \apj, 218, 86
\bibitem[Byrd (1976)]{gB76}  Byrd G.G., 1976, \apj, 208, 688
\bibitem[Byrd (1976,1977,1978)]{gB767778}  
   \phantom{x}
%
\bibitem[Cepa \& Beckman (1988)]{CB88}
   Cepa J., Beckman J.E.,
   1988, \aap, 200,21
%
\bibitem[Cr\'{e}z\'{e} \& Mennessier (1973)]{CM73}
   Cr\'{e}z\'{e}, M., Mennessier, M.O.,
   1973, \aap, 27, 281
%
\bibitem[Corbelli \& Schneider (1997)]{CS1997} 
   Corbelli E.  \&  Schneider S.E.,
   1997, \apj, 479, 244 
%
\bibitem[de Vaucouleurs \etal (1991)]{RC3} 
   de Vaucouleurs G. \etal,
   1991, Volume 1-3, XII, 
   Springer-Verlag, Berlin, Heidelberg, New York
%
\bibitem[Drimmel, Smart \& Lattanzi (2000)]{DSL2000} 
   Drimmel R., Smart R.L. \& Lattanzi M.G.,
   2000, \aap, 354, 67
%
\bibitem[Florido \etal (1991)]{Fea1991} 
   Florido E. \etal,
   1991, \aap, 242, 301
%
\bibitem[Lepine, Mishurov \& Dedikov (2000)]{LMD2000}
   Lepine J.R.D., Mishurov Yu.N., Dedikov S.Yu,
   2000, astro-ph/0001216
%
\bibitem[Lequeux, Dantel-Fort \& Fort (1995)]{LDF1995} 
   Lequeux J., Dantel-Fort M. \& Fort B.,
   1995, \aap, 296, L13 
%
\bibitem[Lin, Yuan \& Shu (1969)]{LYS69}
   Lin, C.C., Yuan, C., Shu, F.H.,
   1969, \apj, 155, 721
%
\bibitem[Magnier \etal (1997)]{Mea97}
   Magnier E.A., Prins S., Augusteijn T., van Paradijs J., Lewin W.H.G.,
   1997, \apj, 326, 442
%
\bibitem[Maucherat \etal (1984)]{Metal1984} 
   Maucherat A.J., Figon P., Dubout-Crillon R. \& Monnet G.,
   1984, \aap, 133, 341 
%
\bibitem[Mishurov \& Zenina (1999)]{MZ99}
   Mishurov, Yu. N., Zenina, I.A.,
   1999, \aap, 341, 81
%
\bibitem[Mishurov et al. (1997)]{MZp97} 
   Mishurov Y.N., Zenina I.A., Dambis A.K., Mel'Nik A.M. \& Rastorguev A.S.,
   1997, \aap, 323, 775 
%
\bibitem[Porcel, Battaner \& Jimenez-Vicente (1997)]{PBJ1997} 
   Porcel C., Battaner E. \& Jimenez-Vicente J.,
   1997, \aap, 322, 103
%
\bibitem[Reshetnikov \& Combes (1998)]{RC1998} 
   Reshetnikov V. \& Combes F.,
   1998, \aap, 337, 9
%
\bibitem[Sanchez-Saavedra, Battaner \& Florido (1990)]{SBF1990} 
   Sanchez-Saavedra M.L., Battaner E. \& Florido E.,
   1990, \mnras, 246, 458
%
\bibitem[Sandage \& Humphreys (1980)]{SH1980} 
   Sandage A. \& Humphreys R.M.,
   1980, \apjl, 236, L1
%
\bibitem[Sato  \& Sawa (1986)]{SS86} 
   Sato N.R. \&  Sawa T.,
   1986, \pasj, 38, 63 
%
\bibitem[The SIM book (1999)]{SIM99}
   The SIM book, 1999
   http://sim.jpl.nasa.gov/library/book.html
%
\bibitem[Tilanus \& Allen (1993)]{TA93} 
   Tilanus R.P.J. \&  Allen R.J.,
   1993, \aap, 274, 707 
%
\bibitem[Unwin (1983)]{U1983} 
   Unwin S.C.,
   1983, \mnras, 205, 773
%
\bibitem[Visser (1980)]{V80} 
   Visser H.C.D.,
   1980, \aap, 88, 159
%
\bibitem[Vogel, Kulkarni \& Scoville (1988)]{VKS88}
   Vogel S.N., Kulkarni S.R. \& Scoville N.Z.,
   1988, \nat, 334, 402 
%
\bibitem[Walterbos \& Kennicutt (1988)]{WK1988} 
   Walterbos R.A.M. \& Kennicutt R.C. Jr.,
   1988, \aap, 198, 61
%
\end{thebibliography}
\end{document}